\documentclass[12pt]{article}
\usepackage{amsmath,amssymb,bm,graphicx}
\begin{document}

\begin{flushright}
\end{flushright}

\vskip 0.5 truecm

\begin{center}
{\Large{\bf Remark on the subtractive renormalization of  quadratically divergent scalar mass}}
\end{center}
\vskip .5 truecm
\centerline{\bf  Kazuo Fujikawa}
\vskip .4 truecm
\centerline {\it Institute of Quantum Science, College of 
Science and Technology}
\centerline {\it Nihon University, Chiyoda-ku, Tokyo 101-8308, 
Japan}
\vskip 0.5 truecm

\makeatletter
\@addtoreset{equation}{section}
\def\theequation{\thesection.\arabic{equation}}
\makeatother

\begin{abstract}
The quadratically divergent scalar mass is subtractively renormalized unlike other divergences which are multiplicatively renormalized. 
We re-examine some technical aspects of the subtractive renormalization, in particular, the mass independent renormalization of massive $\lambda\phi^{4}$ theory with  higher derivative regularization. We then discuss an unconventional scheme to introduce the notion of renormalization point $\mu$ to the subtractive renormalization in a theory defined by
 a large fixed cut-off $M$. The resulting renormalization group equation generally becomes inhomogeneous but it is transformed to be homogeneous. The renormalized scalar mass consists of two components in this scheme, one with the ordinary anomalous dimension and the other which is proportional to the renormalization scale $\mu$. This scheme interpolates between the theory defined by dimensional regularization and the theory with un-subtracted quadratic divergences.
\end{abstract}

\section{Introduction}
The renormalization theory and its application to the Standard Model is very successful~\cite{weinberg1}. The discovery of the Higgs particle at LHC in the predicted mass range will complete the picture of the spontaneous breakdown of gauge symmetry.
A salient feature of the scalar particle such as the Higgs particle is that its mass is generally 
renormalized subtractively unlike other parameters in renormalizable theory. This feature of the scalar mass is also related to the issue of
 naturalness ~\cite{susskind}. There are varying views on naturalness, and we stay neutral on the naturalness issue itself. Rather, we study some 
 technical aspects associated with the multiplicative and subtractive renormalizations.
For definiteness, we define the operational difference between  two renormalizations as follows: For a large but fixed cut-off $M$ and to a finite order in perturbation theory, the renormalized parameter is made small by letting the corresponding bare parameter small in the multiplicative renormalization (as is exemplified by (2.2) and (2.10) below), while this is not the case in the subtractive renormalization (as is shown by the presence of the last term in (2.3) below). Understood in this way, the starting bare Lagrangian controls the multiplicatively renormalized  parameters better. It is our opinion that the naturalness issue in a naive sense would not have been raised if all the parameters in the Standard Model were multiplicatively renormalized, although both renormalizations are perfectly consistent in conventional renormalization theory. 
  
 The subtractive renormalization is associated with the quadratic divergence in scalar field theory such as the $\lambda\phi^{4}$ theory which we study in this paper.  
A way to avoid the quadratic divergence and thus subtractive renormalization is to use the idea of supersymmetry. As is well-known from 
the very beginning of modern supersymmetric field theory, the ultra-violet divergences are improved by supersymmetry~\cite{wess1}. It has been shown that the simplest Wess-Zumino model is renormalized to all orders in perturbation theory by the wave
function renormalization only, either in the component field formulation~\cite{iliopoulos} or in the superfield formulation~\cite{fujikawa}. In the superfield $\phi(x,\theta)$ formulation~\cite{salam}, the supersymmetric  $\lambda\phi^{4}$ theory is reduced to an effective $\phi^{3}(x,\theta)$ theory and the analysis of ultraviolet divergences is much simplified~\cite{sakata}.
Another technical way, which side-steps the issue of  quadratic divergence, is to use the dimensional regularization~\cite{t hooft}. One then encounters no quadratic divergence in the $\lambda\phi^{4}$ theory, and thus avoids the issue of the subtractive renormalization~\cite{collins}. This is obviously a technical solution, but it may have a deeper reason~\cite{bardeen}. In any case, most of the practical calculations in the Standard Model are performed with the dimensional regularization.

The purpose of the present paper is to study the conventional treatment of the quadratic divergence in massive $\lambda\phi^{4}$ theory and the associated subtractive renormalization in further detail. A readable account of the renormalization of $\lambda\phi^{4}$ theory together with past references are found in the monograph by Zinn-Justin\cite{zinn-justin}. Zinn-Justin also gives an interesting argument for the existence of the truly massless $\lambda\phi^{4}$ theory. Our emphasis in the present  paper is on the counter terms for quadratic divergences and their possible implications.
We first re-examine the issue of the quadratic divergence in the so-called mass independent renormalization scheme~\cite{t hooft2, weinberg}. Weinberg  noted the complication caused by the quadratic divergence in scalar theory in his original paper on the mass-independent scheme~\cite{weinberg}. We attempt to deal with this problem in a manner different from the original scheme of Weinberg by introducing a counter term, which is independent of the scalar mass, to subtract the quadratic divergence completely before the conventional multiplicative renormalization. We then define mass-independent renormalization factors in the massive  $\lambda\phi^{4}$ theory. The infrared divergence is related to this analysis, and we give a prescription to avoid the infrared divergence basically working in the framework of the {\em massive} theory. Zinn-Justin~\cite{zinn-justin} gives a different scheme to avoid the infrared divergence. 

In the course of this analysis, we recall that Callan avoided the direct encounter with the quadratic divergence by the mass insertion technique
 in his original treatment of the Callan-Symanzik equation~\cite{callan}. One can thus formally side-step the issue of the quadratic divergence. We note that the consistent elimination of the quadratic divergence by the mass insertion technique is closely related to our specific way of subtracting the quadratic divergence before any multiplicative renormalization. We also note that the procedure of Callan is  related to the classical scaling argument of Bardeen~\cite{bardeen}; both are related to the conformal anomaly.

Another issue we study is the relation of the subtractive renormalization with an inhomogeneous renormalization group equation. As is well-known, the subtractive renormalization generally leads to an inhomogeneous renormalization group equation. An explicit example is the analysis of the $e^{+}e^{-}$ annihilation amplitude in QCD performed by Zee~\cite{zee}. In the ordinary treatment of the quadratic divergence in scalar theory, this inhomogeneous renormalization group equation does not appear. We however note that it is in principle possible to write an inhomogeneous renormalization group equation if one introduces the notion of the renormalization point $\mu$ into the subtraction term of the quadratic divergence; in this formulation we suppose that the magnitude of a large fixed cut-off $M$ has some physical meaning. This leads to an unconventional result that the physical scalar mass generally depends on the renormalization point $\mu$, and each $\mu$ defines a different physical theory for a fixed cut-off $M$. As a result, this scheme interpolates between two different theories, namely, the theory defined by dimensional regularization and the theory with un-subtracted quadratic divergences. 

We also note that our renormalization group equation is similar to 
\begin{eqnarray}
[\Lambda\frac{\partial}{\partial \Lambda}+\beta\frac{\partial}{\partial \lambda}- (\alpha_{1}m^{2}+\alpha_{2}\Lambda^{2})\frac{\partial}{\partial m^{2}}+n\gamma]G(p_{1},p_{2}, ..., p_{n})=0
\end{eqnarray}
which is proposed by Hughes and Liu~\cite{hughes} on the basis of the cut-off parameter $\Lambda$ independence of the Green's function in the Wilsonian renormalization scheme~\cite{kogut, polchinski}. A salient feature of this equation (1.1) is that the mass term generally contains $\Lambda^{2}$ without spoiling the  $\Lambda$ independence of the Green's function. See, for example, Ref.\cite{sonoda} for a recent review of Wilsonian renormalization. In our modified renormalization group equation, $\Lambda$  in (1.1) is replaced by the renormalization scale $\mu$, namely, the parameter $\mu$ plays a role similar, though not identical,  to the cut-off $\Lambda$ in the Wilsonian renormalization. 
 
\section{Mass-independent renormalization}

\subsection{Massive $\lambda\phi^{4}$ theory}

We re-examine the scalar mass renormalization of the $\lambda\phi^{4}$ theory defined in Euclidean space with the metric $g_{\mu\nu}=(1,1,1,1)$. To specify a better defined theory, one may start with  
\begin{eqnarray}
{\cal L}&=&-\frac{1}{2}\phi_{0}(x)[-\Box+m_{0}^{2}](\frac{-\Box+M^{2}}
{M^{2}})^{2}\phi_{0}(x)-\frac{1}{4!}\lambda_{0}\phi_{0}(x)^{4}
\end{eqnarray}
and renormalize the theory multiplicatively by  
\begin{eqnarray}
&&\phi_{0}(x)=\sqrt{Z( \lambda_{0}, M, m_{0})}\phi(x),\nonumber\\
&&m_{0}^{2}=\frac{Z_{m}(\lambda_{0}, M, m_{0})}{Z(\lambda_{0},  M, m_{0})}m^{2},\nonumber\\
&&\lambda_{0}=\frac{Z_{\lambda}(\lambda_{0},  M, m_{0})}{Z^{2}(\lambda_{0},  M, m_{0})}\lambda
\end{eqnarray}
in the bare perturbation theory.
The parameter $M$ provides the ultraviolet cut-off and a large mass proportional to $M^{2}$ is induced in this scheme.

To avoid the large induced mass, one may next specify  the "bare" Lagrangian by 
\begin{eqnarray}
{\cal L}&=&-\frac{1}{2}\phi_{0}(x)[-\Box+m_{0}^{2}](\frac{-\Box+M^{2}}
{M^{2}})^{2}\phi_{0}(x)-\frac{1}{4!}\lambda_{0}\phi_{0}(x)^{4}\nonumber
\\
&&+\frac{1}{2}\Delta_{sub}(\lambda_{0},M^{2})\phi_{0}(x)^{2}
\end{eqnarray}
where $\Delta_{sub}(\lambda_{0},M^{2})$ is chosen such that all the induced mass terms proportional to $M^{2}$ are completely subtracted.
Our definition of the bare mass $m^{2}_{0}$ differs from the common definition~\cite{zinn-justin} but in accord with the dimensional regularization. In this scheme, the free propagator is given by 
\begin{eqnarray}
\int d^{4}xe^{ipx}\langle T \phi_{0}(x)\phi_{0}(0)\rangle=\frac{1}{p^{2}+m_{0}^{2}}(\frac{M^{2}}{p^{2}+M^{2}})^{2}.
\end{eqnarray}
A salient feature of (2.3) is the choice of the subtraction term which is independent of $m_{0}^{2}$
\begin{eqnarray}
m_{0}\frac{d}{dm_{0}}\Delta_{sub}(\lambda_{0},M^{2})=0.
\end{eqnarray}
To ensure this property, it is important to subtract all the quadratic divergences up to any finite order in perturbation theory {\em before}
any multiplicative renormalization. The property (2.5) itself is apparently well-known~\cite{zinn-justin} though not emphasized, We re-examine this property in some detail since it is crucial for our entire analysis. As an 
illustration, we examine the direct evaluation of the two-loop mass term for the Lagrangian (2.3) in Appendix and show that the term proportional to $\lambda_{0}^{2}M^{2}\ln(M^{2}/m_{0}^{2})$, which spoils our assumption (2.5), does not appear in the quadratic divergence. After this subtraction of the quadratic divergence, the remaining part of the self-energy diagram is at most logarithmically divergent and the complications of the quadratic  divergence are avoided. We describe later how this subtraction of quadratic divergences generally works.

To support the assumption in (2.5), we here give two general arguments. 
If the property (2.5) does not hold, the conventional form of the Callan-Symanzik equation~\cite{callan} would contain an extra inhomogeneous term 
coming from $m_{0}\frac{d}{dm_{0}}\Delta_{sub}$. The simplification 
of the renormalization analysis of the $\phi^{4}$ theory by sidestepping the overlapping divergence on the basis of the Callan-Symanzik equation~\cite{callan}, which crucially depends on the choice (2.5), is then spoiled. Secondly, the subtraction procedure of the quadratic divergence in (2.3) is analogous to the dimensional regularization where the quadratic divergence is completely subtracted
before the conventional multiplicative renormalization; in fact, the consistent operation of the dimensional regularization~\cite{collins}
without spoiling the physical contents (and without encountering the quadratic divergence) suggests that the choice $\Delta_{sub}(\lambda_{0},M^{2})$ is possible.

The same result is realized by rewriting (2.3) as 
\begin{eqnarray}
{\cal L}&=&-\frac{1}{2}\phi_{0}(x)[-\Box(\frac{-\Box+M^{2}}
{M^{2}})^{2}]\phi_{0}(x)-\frac{1}{2}m_{0}^{2}\phi_{0}(x)\phi_{0}(x)-\frac{1}{4!}\lambda_{0}\phi_{0}(x)^{4}\nonumber
\\
&&+\frac{1}{2}\Delta_{sub}(\lambda_{0},M^{2})\phi_{0}(x)^{2}
\end{eqnarray}
and treating the mass term as a part of the interaction. Formally, (2.3) and (2.6) define an identical theory. This mass independent renormalization scheme, which greatly simplifies the analysis of the renormalization group equation, was introduced in~\cite{t hooft2} 
and~\cite{weinberg}. The formulation of 't Hooft~\cite{t hooft2}, which is based on the dimensional regularization, does not encounter the quadratic divergence and thus we do not need the last term in (2.6); in fact one can directly work with the massive theory (2.3) without the higher derivative regularization. The formulation of Weinberg~\cite{weinberg} is similar to the formulation in (2.6). The potential complication of the mass independent scheme
due to the quadratically divergent scalar mass was noted by Weinberg. On the basis of the detailed analysis in the dimensional regularization by Collins~\cite{collins} and also the analysis described below, we assume that the systematic subtraction of quadratic divergences works both in (2.3) and (2.6). This is also in accord with the previous analysis by Zinn-Justin~\cite{zinn-justin}. To cope with infrared divergences in (2.6), however, we operate in a scheme different from the original scheme of Weinberg
 and discuss how to define mass-independent renormalization factors  basically in the {\em massive} perturbation theory defined by (2.3) and (2.4).
 
\subsection{ Analysis of ultraviolet divergence}

We now sketch how the systematic subtraction of the quadratic divergence works. This analysis of the quadratic divergence is more transparent in the above mass independent  bare perturbation theory
defined by (2.6). We thus start with the propagator given by 
\begin{eqnarray}
\int d^{4}xe^{ipx}\langle T \phi_{0}(x)\phi_{0}(0)\rangle=\frac{1}{p^{2}}(\frac{M^{2}}{p^{2}+M^{2}})^{2}.
\end{eqnarray}
We first note that the "primitive" quadratically divergent diagram which does not contain any quadratically divergent sub-diagrams is infrared finite. Here the quadratically divergent diagrams mean the diagrams 
whose superficial degree of divergence is two.
Some of the examples are given in Fig.1.

\begin{center}
\includegraphics[scale=0.4,clip]{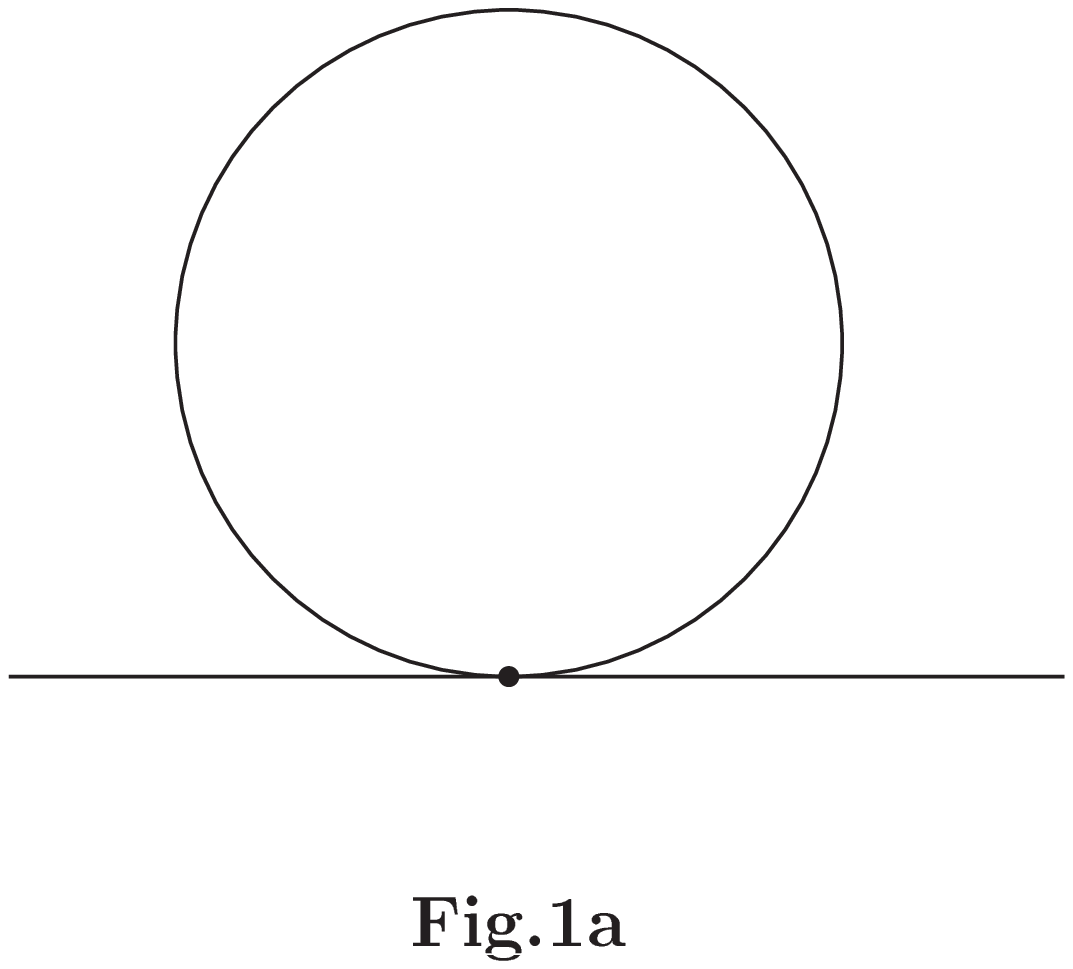}
\hspace{3mm}
\includegraphics[scale=0.4,clip]{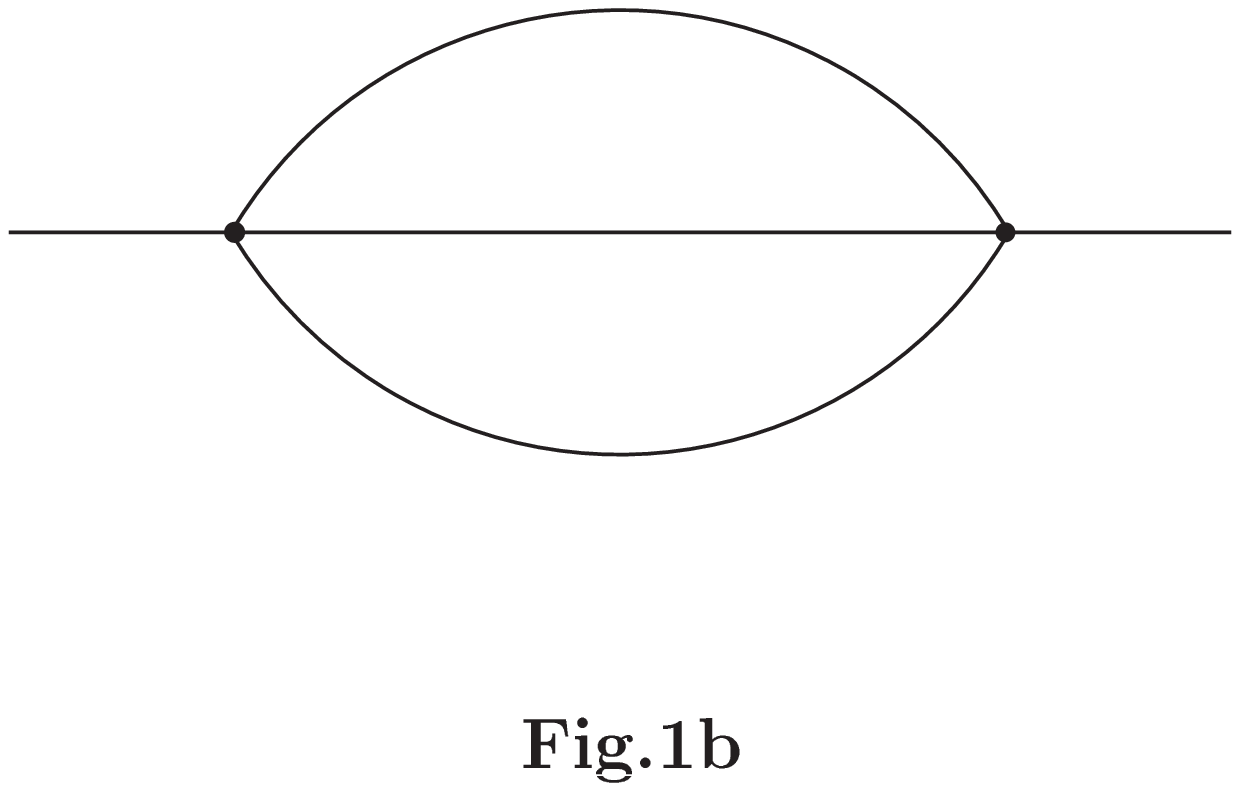}
\end{center}

\vspace{2mm}

\begin{center}
\includegraphics[scale=0.4,clip]{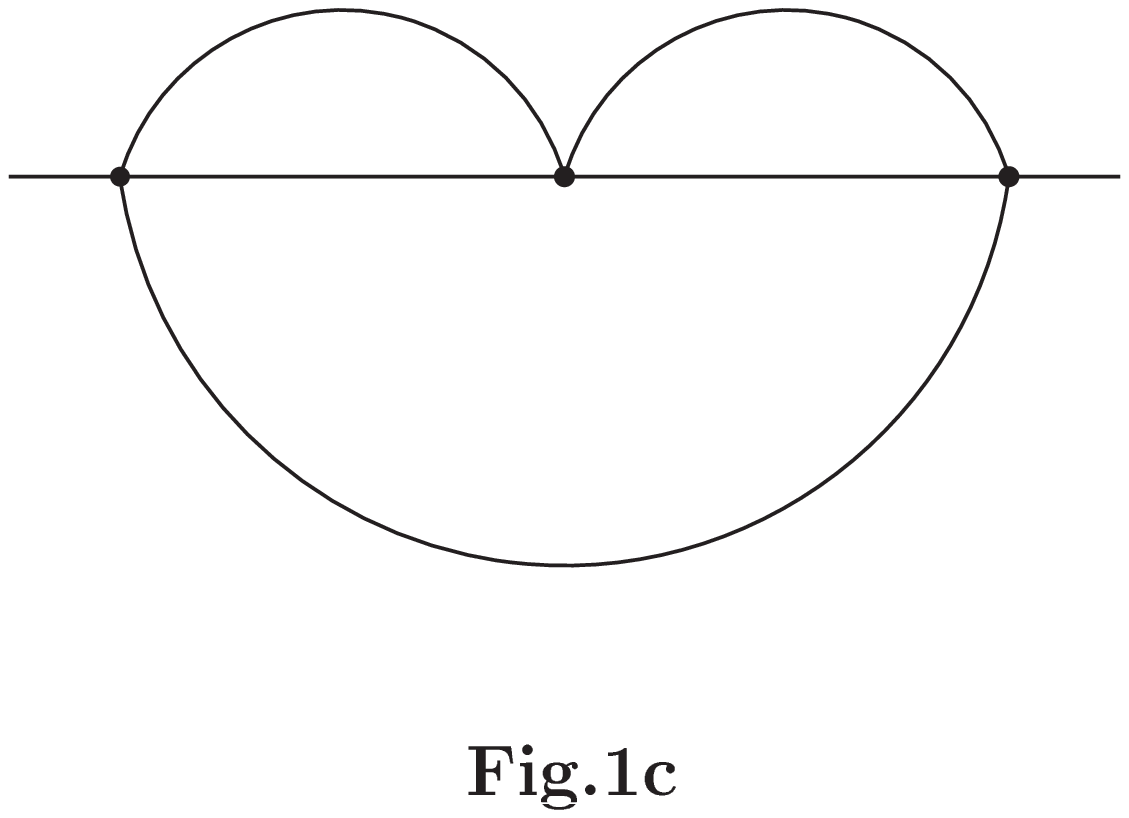}
\hspace{3mm}
\includegraphics[scale=0.4,clip]{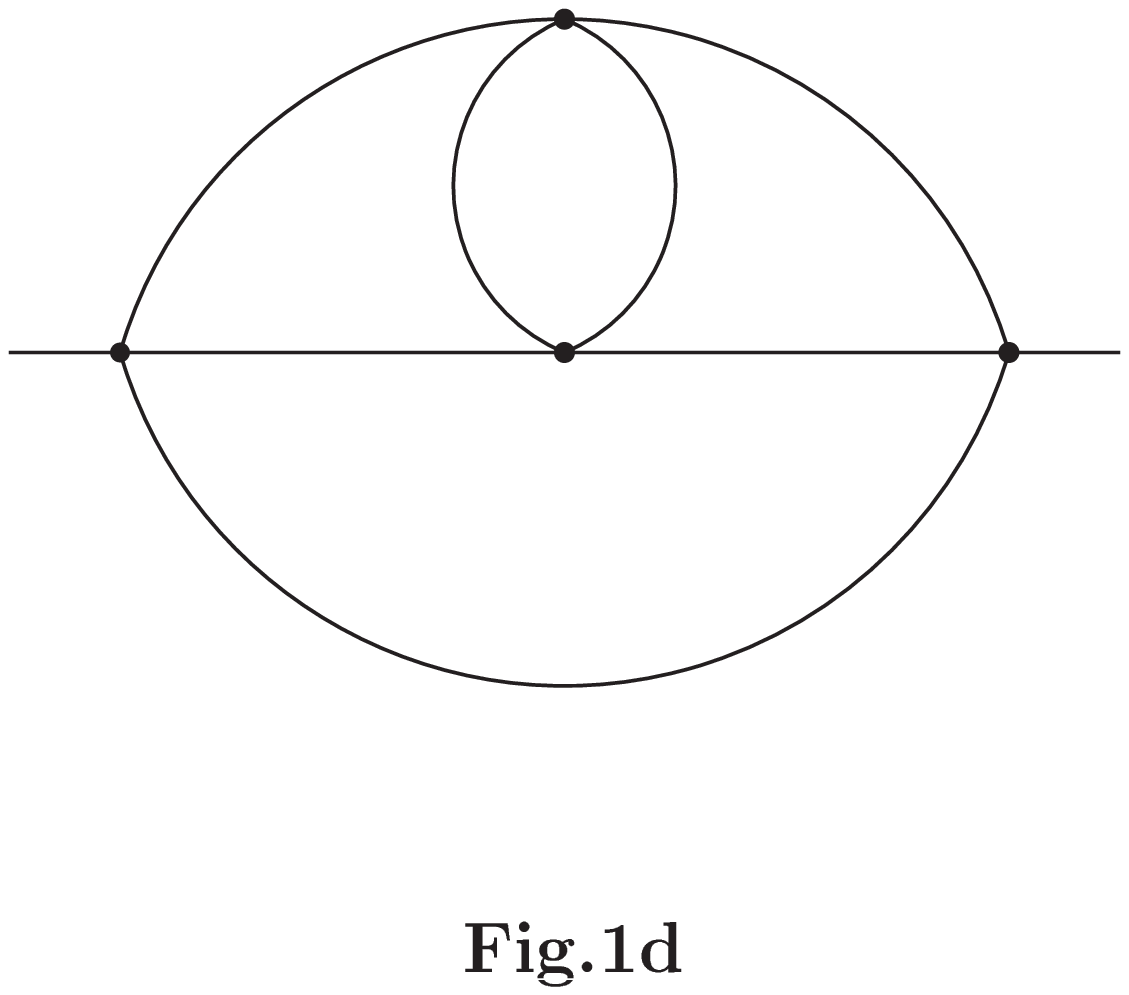}
\end{center}

Fig.1a is explicitly dealt with in Section 4. The infrared 
finiteness of Fig.1b is shown in Appendix. One can confirm the infrared finiteness of Figs.1c and 1d by a power counting argument.   
One may then evaluate any of these self-energy diagrams in Fig.1 to obtain
\begin{eqnarray}
\Sigma(p^{2}, M^{2},\lambda_{0})
\end{eqnarray}
by using the propagator (2.7) in the mass independent scheme. We then subtract the quadratic divergence by 
\begin{eqnarray}
\tilde{\Sigma}(p^{2}, M^{2},\lambda_{0})&=&\Sigma(p^{2}, M^{2},\lambda_{0})-\Sigma(0, M^{2},\lambda_{0})\nonumber\\
&\equiv&p^{2}A(p^{2}/M^{2},\lambda_{0}).
\end{eqnarray}
The quantity $\tilde{\Sigma}(p^{2},M^{2},\lambda_{0})$ thus defined is logarithmically divergent in general in the ultraviolet for large $M$, and $A(p^{2}/M^{2},\lambda_{0})$  generally contains the (logarithmic) infrared singularity at $p^{2}=0$ . The constant $\Sigma(0, M^{2},\lambda_{0})$ constitutes a part of the counter term in (2.6)
in the corresponding order in  perturbation theory. The quantity $\tilde{\Sigma}(p^{2},M^{2},\lambda_{0})$ in (2.9) identically vanishes for  
massless tadpole-type diagrams such as Fig.1a, and thus our prescription resembles the  normal ordering prescription. But the massive tadpole-type diagrams are not eliminated by our prescription 
in accord with the prescription in the dimensional regularization\cite{t hooft}.

When one analyzes those quadratically divergent diagrams  which contain
one or more "primitive" quadratically divergent sub-diagrams, one needs to take care of the possible infrared singularity. Some examples of these diagrams are shown in Fig.2. 

\vspace{2mm}

\begin{center}
\includegraphics[scale=0.6,clip]{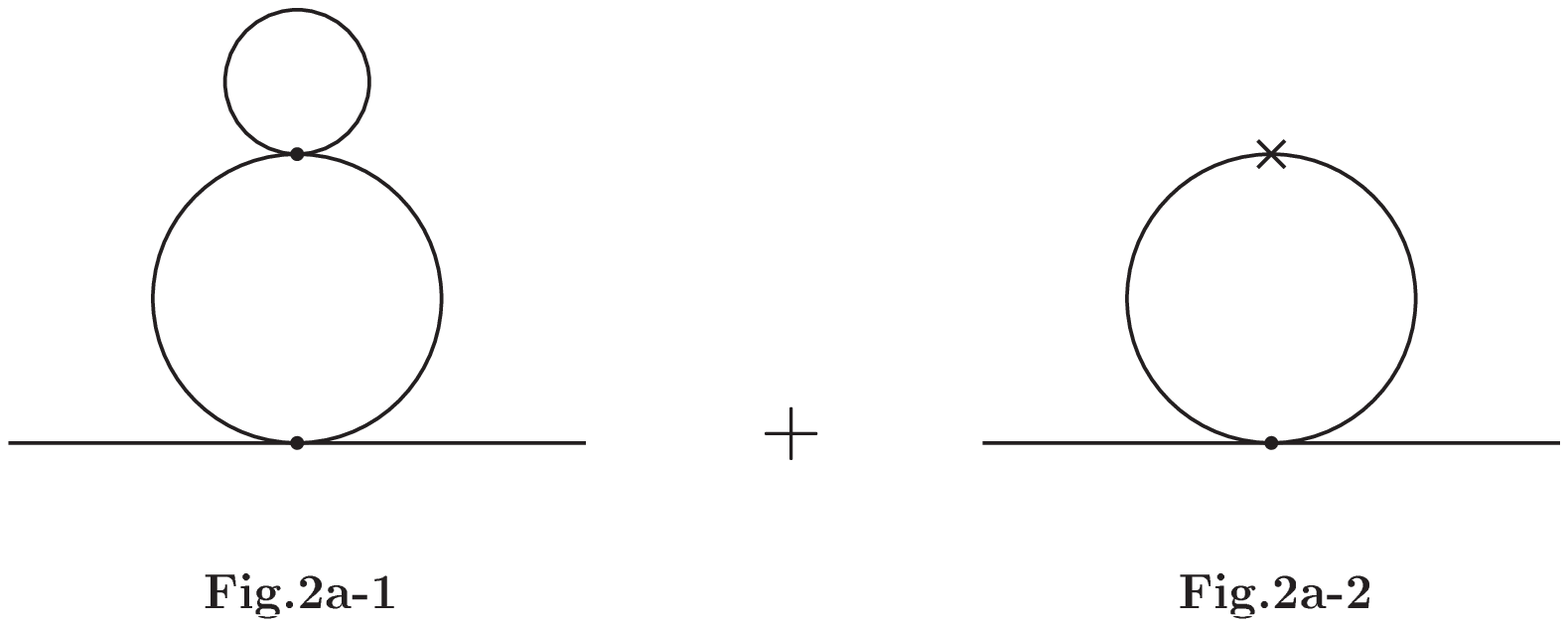}
\end{center}

\vspace{2mm}

\begin{center}
\includegraphics[scale=0.6,clip]{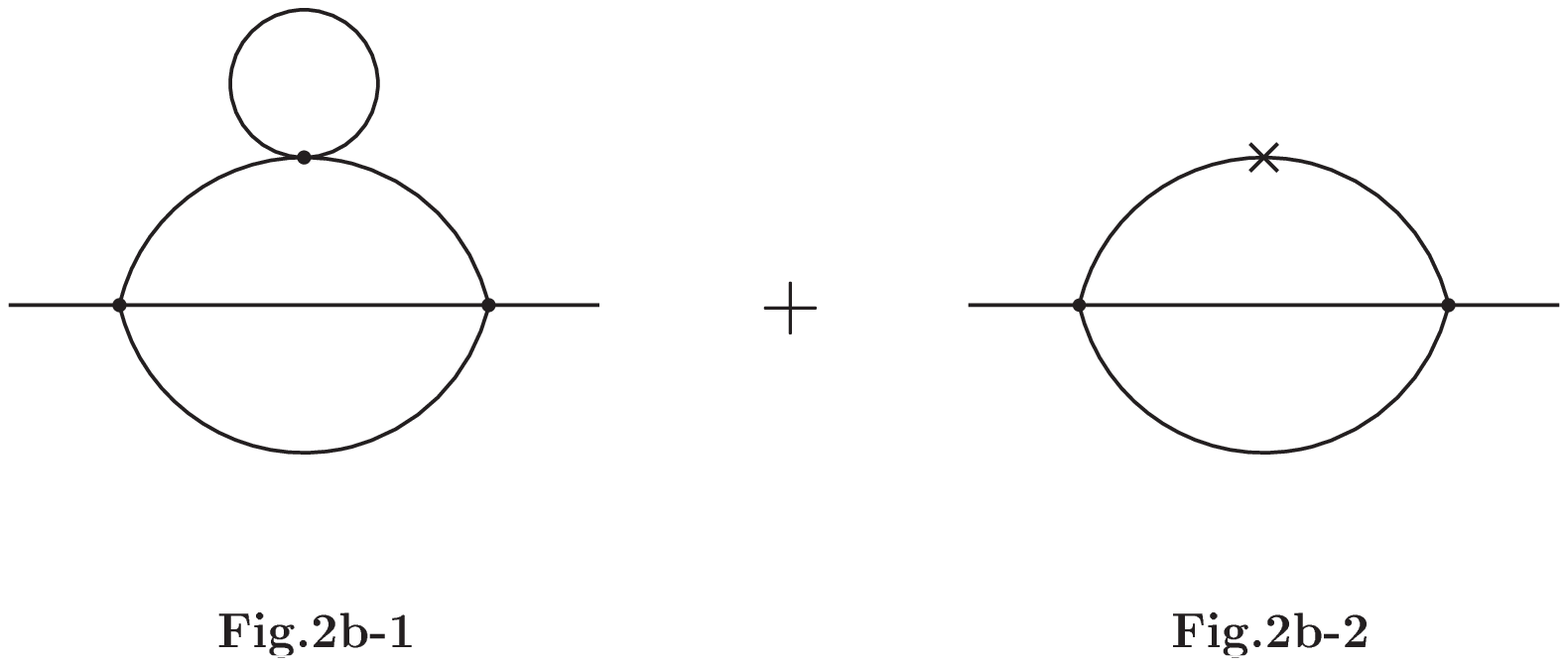}
\end{center}

\vspace{2mm}

\begin{center}
\includegraphics[scale=0.6,clip]{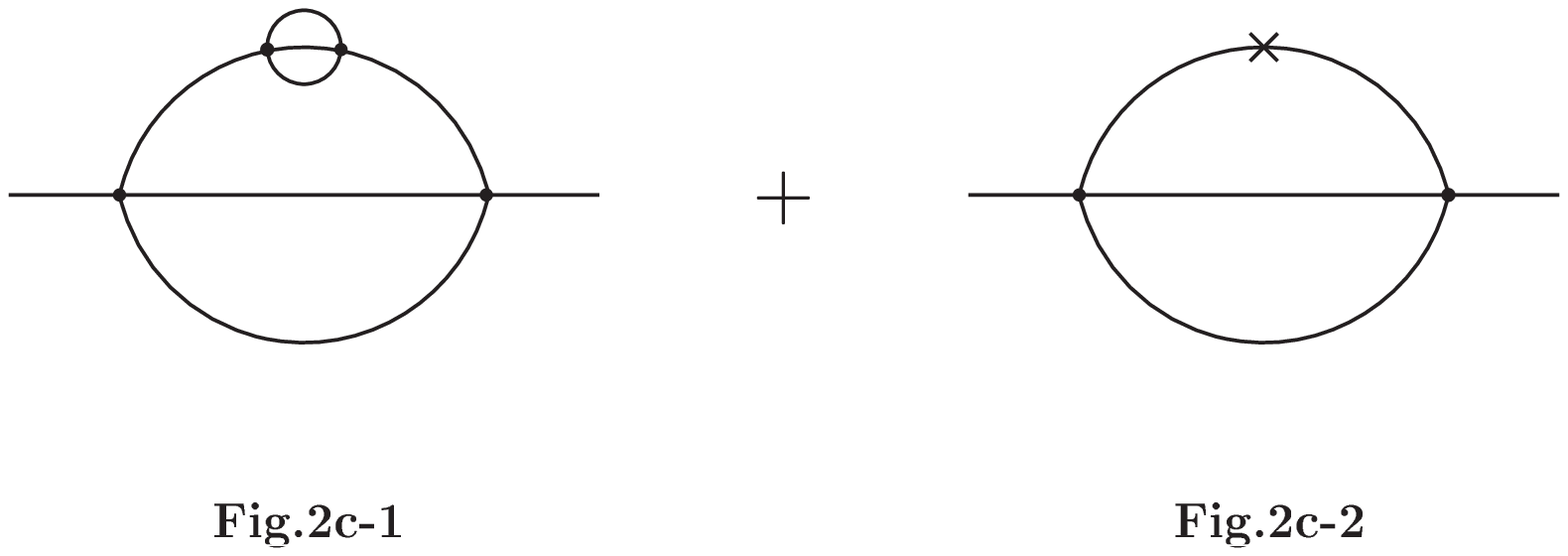}
\end{center}

One can confirm that Figs.2a-1, 2b-1 and 2c-1 are all infrared divergent for the massless propagator in (2.7), and those infrared divergences are not controlled by the external Euclidean momentum flowing into the diagrams. But when one combines Fig.2a-1 with  Fig.2a-2, for example, the infrared divergence
is cancelled. Here the cross in Fig.2a-2 stands for $-\Sigma(0, M^{2},\lambda_{0})$ corresponding to Fig.1a. Similarly, the combinations of 
Fig.2b-1 and Fig.2b-2 or Fig.2c-1 and Fig.2c-2 are infrared finite if one 
uses $-\Sigma(0, M^{2},\lambda_{0})$ corresponding to Fig.1a or Fig.1b,
respectively. In those combinations, one effectively replaces $\Sigma(p^{2}, M^{2},\lambda_{0})$ by $\tilde{\Sigma}(p^{2}, M^{2},\lambda_{0})$
in (2.9) for self-energy sub-diagrams and thus the diagrams with massless tadpole insertions, such as Fig.2a and Fig.2b, are completely eliminated
in accord with the dimensional regularization\cite{t hooft}. 

One then defines the over-all subtraction constants of quadratic divergences by setting $p^{2}=0$ in those (surviving) infrared-free combinations. The subtraction constant $\Delta_{sub}(\lambda_{0},M^{2})$ in (2.6) is given by the sum of all these subtraction constants in each given order in perturbation theory. This subtraction of quadratic divergences works for the massive perturbation theory in (2.3) also as is explained in some detail later. The present prescription is thus close to that in dimensional regularization, and in fact one may regard our prescription as a Lagrangian implementation of  dimensional 
regularization when it comes to the elimination of quadratic divergences.

By this procedure, one can generate the self-energy amplitudes order by order in the bare perturbation theory which are free of quadratic divergences. One then applies the general renormalization procedure in the bare perturbation theory to those self-energy amplitudes  at the off-shell point $p^{2}=\mu^{2}$ to define the wave function renormalization factor~\cite{weinberg}; the quantity $\tilde{\Sigma}(p^{2}, M^{2},\lambda_{0})$ in (2.9) generally contains both coupling constant and wave function renormalization factors.  The mass insertion diagrams or the four-point proper vertices for the Lagrangian (2.6), which do not directly induce the quadratic divergence, are handled after removing the possible quadratic divergences in sub-diagrams by the procedure described above. 

In practice, however, one needs a careful treatment of infrared singularities in those logarithmically (ultraviolet) divergent diagrams. In particular, mass insertion diagrams contain infrared divergences which are not controlled by external momentum flowing into the diagrams. See, for example, mass insertion diagrams in Fig.3 which are infrared divergent.

\begin{center}
\includegraphics[clip,scale=0.45]{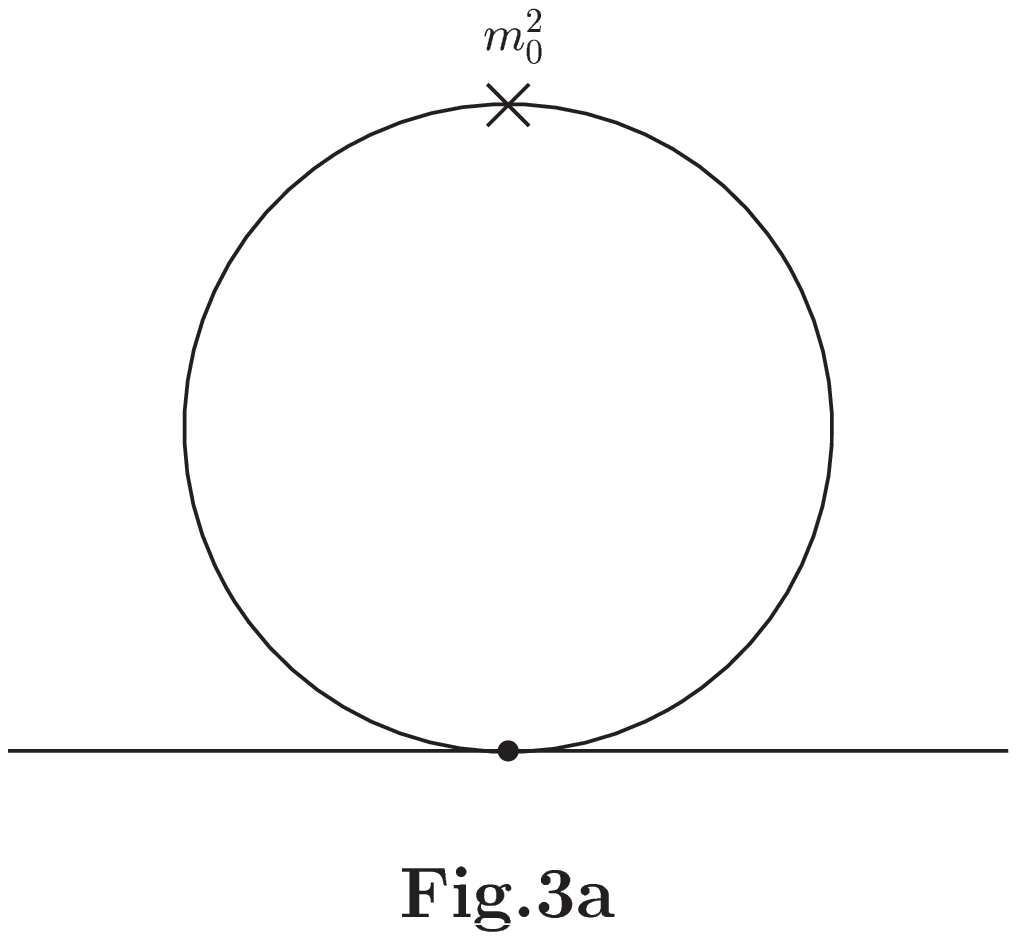}
\hspace{5mm}
\includegraphics[clip,scale=0.45]{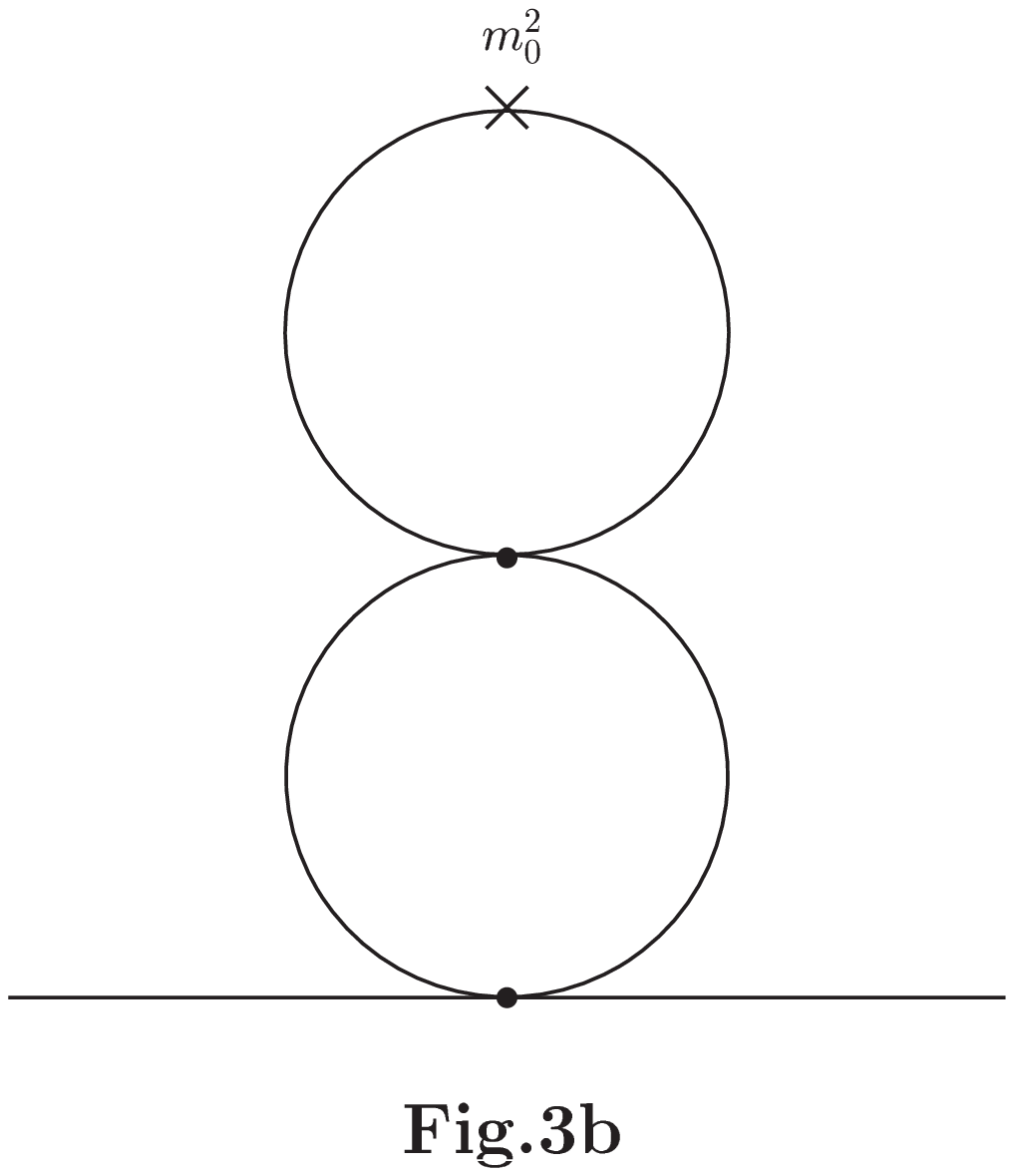}
\hspace{5mm}
\end{center}

\vspace{3mm}

\begin{center}
\includegraphics[clip,scale=0.45]{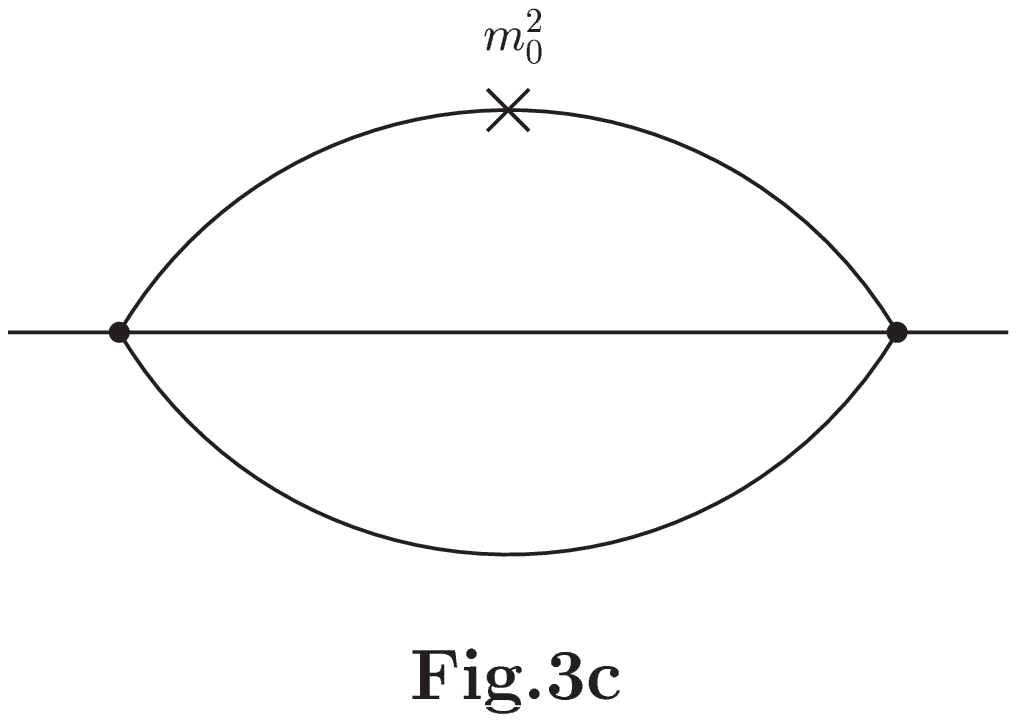}
\end{center}
If one ignores the infrared divergence for a moment, the ultraviolet logarithmic divergence in Fig.3a is absorbed by the order $\lambda_{0}$ renormalization of the tree level mass term. The logarithmic divergences of lower loop sub-diagrams in 
Fig.3b and Fig.3c together are absorbed by the order $\lambda^{2}_{0}$
renormalization of the coupling constant in Fig.3a. In fact, one can 
generate Fig.3a$\sim$Fig.3c by first drawing the tree level and 
one loop four-point diagrams and then adding the mass term as an interaction. The remaining divergences coming from the upper loop diagrams 
in Fig.3b and Fig.3c are then partly absorbed by the order  $\lambda_{0}$
mass renormalization in Fig.3a and the rest of the logarithmic divergence is absorbed by the order $\lambda^{2}_{0}$ renormalization of the tree level mass term.

\subsection{Analysis of infrared divergence}

In reality, one needs to take care of the infrared divergence in those 
diagrams such as in Fig.3. To deal with the infrared divergence we go back to the massive perturbation defined by (2.3) and (2.4) which is manifestly free of infrared divergences, although the isolation of the quadratic divergence is more transparent in the mass independent scheme in (2.6). We here depart from the original formulation of Weinberg\cite{weinberg} and attempt to define the mass independent renormalization factors in the form 
\begin{eqnarray}
&&\phi_{0}(x)=\sqrt{Z( \lambda_{0}, M, \mu)}\phi(x),\nonumber\\
&&m_{0}^{2}=\frac{Z_{m}(\lambda_{0}, M, \mu)}{Z(\lambda_{0},  M, \mu)}m^{2},\nonumber\\
&&\lambda_{0}=\frac{Z_{\lambda}(\lambda_{0},  M, \mu)}{Z^{2}(\lambda_{0},  M, \mu)}\lambda
\end{eqnarray}
for the massive theory defined by (2.3) and (2.4). In our later analysis of the renormalization group equation, the mass independent renormalization plays an essential role.

We start with the coupling constant renormalization, which is logarithmically divergent in the ultraviolet and infrared divergent in the massless limit. For those logarithmically divergent diagrams, we make the replacement 
\begin{eqnarray}
\frac{1}{l^{2}+m_{0}^{2}}(\frac{M^{2}}{l^{2}+M^{2}})^{2}&=&
\frac{1}{l^{2}+\mu^{2}}(\frac{M^{2}}{l^{2}+M^{2}})^{2}\nonumber\\
&+&\frac{\mu^{2}-m_{0}^{2}}{(l^{2}+m_{0}^{2})(l^{2}+\mu^{2})}(\frac{M^{2}}{l^{2}+M^{2}})^{2}
\end{eqnarray}
in the propagator (2.4). For the vertex correction in the massive theory (in the one-loop level, for example), we have 
\begin{eqnarray}
&&\frac{\lambda_{0}^{2}}{2}\int \frac{d^{4}l}{(2\pi)^{4}}\frac{1}{(l+p)^{2}+m_{0}^{2}}\frac{1}{l^{2}+m_{0}^{2}}\nonumber\\
&&=\frac{\lambda_{0}^{2}}{2}\int \frac{d^{4}l}{(2\pi)^{4}}\frac{1}{(l+p)^{2}+\mu^{2}}\frac{1}{l^{2}+\mu^{2}}+\ 
{\rm finite\  terms}\nonumber\\
&&=\frac{\lambda_{0}^{2}}{2}\int \frac{d^{4}l}{(2\pi)^{4}}\frac{1}{l^{2}+\mu^{2}}\frac{1}{l^{2}+\mu^{2}}\nonumber\\
&&+\frac{\lambda_{0}^{2}}{2}\int \frac{d^{4}l}{(2\pi)^{4}}[\frac{1}{(l+p)^{2}+\mu^{2}}-\frac{1}{l^{2}+\mu^{2}}]\frac{1}{l^{2}+\mu^{2}}\nonumber\\
\nonumber\\
&&+\ 
{\rm finite\  terms}
\end{eqnarray}
where $p$ stands for the external momentum.
 Here and in the rest of this section, the regularization factor $M^{4}/(l^{2}+M^{2})^{2}$ for each propagator is implicit. The first term on the right-hand side of (2.12) is logarithmically divergent and absorbed by the coupling constant renormalization with the parameter $\mu$ specifying the renormalization mass scale. The rest of the terms on the right-hand side of (2.12) give finite contributions. This procedure is free of infrared divergences.
 
 Since the logarithmically divergent mass term arises from the mass insertion to the quadratically divergent self-energy diagram in massless perturbation theory and that only the single mass insertion diagrams are logarithmically divergent, we replace the propagator
 in "primitive" quadratically divergent diagrams in the {\em massive} perturbation theory as
\begin{eqnarray}
\frac{1}{k^{2}+m_{0}^{2}}&=&\frac{1}{k^{2}}-m_{0}^{2}\frac{1}{k^{2}(k^{2}+m_{0}^{2})}\\
&=&\frac{1}{k^{2}}-\frac{m_{0}^{2}}{(k^{2}+\mu^{2})^{2}}-\frac{m_{0}^{2}\mu^{2}}{k^{2}(k^{2}+\mu^{2})^{2}}
+\frac{m_{0}^{2}(m_{0}^{2}-\mu^{2})}{k^{2}(k^{2}+\mu^{2})(k^{2}+m_{0}^{2})}\nonumber.
\end{eqnarray}
The first term on the right-hand side of (2.13), which defines the propagator in massless theory, gives rise to the quadratically divergent in the ultraviolet but infrared finite contributions, as was already analyzed. The second term corresponds to the mass insertion such as in Fig.3a$\sim$Fig.3c which are logarithmically divergent in the ultraviolet but infrared finite. These diagrams are renormalized at the mass scale $\mu^{2}$. The last two terms in (2.13), which give both ultraviolet and infrared finite contributions, correspond to the adjustment factor to recover the result of the (infrared divergence free) massive perturbation theory defined by the Lagrangian (2.3). Since only the logarithmically (ultraviolet-) divergent diagrams or sub-diagrams, which are linear in $m_{0}^{2}$,
are important for the mass renormalization, one can define the mass independent renormalization factor by this procedure without encountering infrared divergences.

The diagram in Fig.3a ia explicitly analyzed by using (2.13) in Section 4. The diagram in Fig.3b is a combination of the vertex renormalization (2.12) and Fig.3a. We here briefly illustrate the treatment of Fig.3c which contains the "primitive" quadratic divergence. After the replacement (2.13) and the subtraction of the overall quadratic divergence at the vanishing external momentum in the manner described already, we have for Fig.3c,
\begin{eqnarray}
&&\frac{\lambda_{0}^{2}}{3!}\int \frac{d^{4}k}{(2\pi)^{4}}\frac{d^{4}l}{(2\pi)^{4}}\{[\frac{1}{(k+p)^{2}}-\frac{1}{k^{2}}]\frac{1}{(k+l)^{2}l^{2}}\nonumber\\
&&\hspace{2cm} -\frac{3m_{0}^{2}}{[(k+p)^{2}+\mu^{2}]^{2}}\frac{1}{(k+l)^{2}l^{2}}\nonumber\\
&&\hspace{2cm} -\frac{3m_{0}^{2}\mu^{2}}{(k+p)^{2}[(k+p)^{2}+\mu^{2}]^{2}}\frac{1}{(k+l)^{2}l^{2}}\\
&&\hspace{2cm}+\frac{3m_{0}^{2}(m_{0}^{2}-\mu^{2})}{(k+p)^{2}[(k+p)^{2}+\mu^{2}][(k+p)^{2}+m_{0}^{2}]}\frac{1}{(k+l)^{2}l^{2}}\}.\nonumber
\end{eqnarray}
where we ignored finite terms, and $p$ stands for the external momentum.

 All the terms in (2.14) are infrared finite, and all the integrals are well-defined due to the implicit $M^{4}/(k^{2}+M^{2})^{2}$ for all the propagators. The first term in (2.14) contains the wave function renormalization, the second term contains the coupling constant and mass renormalization, and the last two terms contain the coupling constant renormalization in the present bare perturbation theory. To define the coupling constant renormalization in conformity with (2.12) we make a replacement 
\begin{eqnarray}
\frac{1}{l^{2}}=\frac{1}{l^{2}+\mu^{2}}+\frac{\mu^{2}}{l^{2}(l^{2}+\mu^{2})},
\end{eqnarray}
namely,
\begin{eqnarray}
&&\frac{1}{(k+l)^{2}l^{2}}=\frac{1}{(k+l)^{2}+\mu^{2}}\frac{1}{l^{2}+\mu^{2}}
+\frac{1}{(k+l)^{2}+\mu^{2}}\frac{\mu^{2}}{l^{2}(l^{2}+\mu^{2})}\\
&&+\frac{\mu^{2}}{(k+l)^{2}[(k+l)^{2}+\mu^{2}]}\frac{1}{l^{2}+\mu^{2}}+\frac{\mu^{2}}{(k+l)^{2}[(k+l)^{2}+\mu^{2}]}\frac{\mu^{2}}{l^{2}(l^{2}+\mu^{2})}\nonumber
\end{eqnarray}
in (2.14).
All the four terms in (2.16) give infrared finite contributions when inserted into (2.14), and only the first term in (2.16) gives an ultraviolet divergent  contribution in (2.14). One can thus handle the coupling constant renormalization in accord with (2.12) by retaining only the first term in (2.16) in the last three logarithmically divergent terms in (2.14).  
 
The first term in (2.14) is slightly more involved. By retaining only the first term in (2.16), we have   
\begin{eqnarray}
&&\frac{\lambda_{0}^{2}}{3!}\int \frac{d^{4}k}{(2\pi)^{4}}\frac{d^{4}l}{(2\pi)^{4}}\left(\frac{1}{(k+p)^{2}}-\frac{1}{k^{2}}\right)
\frac{1}{(k+l)^{2}+\mu^{2}}\frac{1}{l^{2}+\mu^{2}}\nonumber\\
&&=\frac{\lambda_{0}^{2}}{3!}\int \frac{d^{4}k}{(2\pi)^{4}}\frac{d^{4}l}{(2\pi)^{4}}\left(\frac{1}{(k+p)^{2}}-\frac{1}{k^{2}}\right)
[\frac{1}{(k+l)^{2}+\mu^{2}}\frac{1}{l^{2}+\mu^{2}}-\frac{1}{(l^{2}+\mu^{2})^{2}}]\nonumber\\
&&\equiv p^{2}A(p^{2},\mu^{2},M^{2})\nonumber\\
&&=p^{2}A(\mu^{2},\mu^{2},M^{2})+p^{2}[A(p^{2},\mu^{2},M^{2})-A(\mu^{2},\mu^{2},M^{2})].
\end{eqnarray}
The equality in the second line in (2.17) shows that the extra subtraction, which formally appears to be the coupling constant renormalization, does not contribute to the present calculation of the wave function renormalization where the quadratic divergence is subtracted at the vanishing external momentum. We thus do not need the subtraction of the potential logarithmic divergence, which formally appears to be the coupling constant renormalization, in the  example in (2.17) unlike the last 3 terms in (2.14); in fact, the number of potential subtraction terms does not match by a factor of 3 for the first term in (2.14). We can thus directly deal with the first term in (2.14), which amounts to use $A(p^{2},0,M^{2})$ in (2.17). The logarithmically divergent wave function renormalization factor $A(\mu^{2},\mu^{2},M^{2})$ is then replaced by $A(\mu^{2},0,M^{2})$, which differ by a finite renormalization. Note that the condition $p^{2}=\mu^{2}$ is added by hand for the wave function renormalization.

The general strategy is now clear. When one considers the mass insertion to a quadratically divergent diagram, only the single mass insertion  is important for the mass renormalization. The single mass insertion to any "primitive" quadratically divergent diagrams is treated as in (2.13) and (2.14) by using (2.15) above. But when one considers a larger quadratically divergent diagram which contains any "primitive" quadratically divergent sub-diarams, 
to which a single mass is inserted by means of (2.13), one encounters the infrared divergence in general if one uses the massless propagator. See, for example, Fig.3b. When one inserts a mass term to "primitive" quadratically divergent sub-diagrams by means of (2.13), which reduces the overall quadratic divergence to the logarithmic divergence, one needs to apply the replacement (2.11) for the propagators {\em outside} any "primitive" quadratically divergent sub-diagrams contained in the larger diagram. 
 
This is the sketch of our calculational procedure of the mass independent multiplicative renormalization factors in the massive perturbation theory. Our analysis is concerned with the elimination of ultraviolet divergences in a mass independent manner, and one generally has to perform additional finite renormalization to specify the precise renormalization conditions~\cite{zinn-justin, weinberg}.

\subsection{Comparison with past analyses}

If one accepts the elimination of the quadratic divergence in (2.3) with
the renormalization factors in (2.2), namely, with $\mu^{2}=m_{0}^{2}$, one may consider a Lagrangian
\begin{eqnarray}
{\cal L}&=&-\frac{1}{2}\phi_{0}(x)[-\Box+\mu^{2}](\frac{-\Box+M^{2}}
{M^{2}})^{2}\phi_{0}(x)-\frac{1}{4!}\lambda_{0}\phi_{0}(x)^{4}
\\
&&-\frac{1}{2}(m_{0}^{2}-\mu^{2})\phi_{0}(x)(\frac{-\Box+M^{2}}{M^{2}})^{2}\phi_{0}(x)+\frac{1}{2}\Delta_{sub}(\lambda_{0},M^{2})\phi_{0}(x)^{2}
\nonumber
\end{eqnarray}
which is {\em identical} to (2.3). One may then regard the 3rd mass term as a 
part of interaction and thus generalize the scheme of Weinberg~\cite{weinberg} in a manner which is free of  infrared divergences for $\mu^{2}\neq 0$. 
The arbitrary
parameter $\mu$ plays a role of the renormalization point, and in fact (2.18) is essentially equivalent to our scheme described above.
The free propagator is given by 
\begin{eqnarray}
\int d^{4}xe^{ipx}\langle T \phi_{0}(x)\phi_{0}(0)\rangle=\frac{1}{p^{2}+\mu^{2}}(\frac{M^{2}}{p^{2}+M^{2}})^{2},
\end{eqnarray}
and  a single  insertion of the mass term to each propagator in a quadratically divergent diagram (to be precise, starting with sub-diagrams if it contains quadratically divergent sub-diagrams), which is relevant to the mass renormalization, amounts to the replacement of the propagator in the diagram
\begin{eqnarray}
\frac{1}{p^{2}+\mu^{2}}- \frac{1}{p^{2}+\mu^{2}}(m_{0}^{2}-\mu^{2})\frac{1}{p^{2}+\mu^{2}}&=&\frac{1}{p^{2}}-\frac{1}{p^{2}+\mu^{2}}m_{0}^{2}\frac{1}{p^{2}+\mu^{2}}\nonumber\\
&-& \frac{1}{p^{2}}\frac{\mu^{4}}{(p^{2}+\mu^{2})^{2}}.
\end{eqnarray}
 We then follow the procedure with (2.13) by maintaining infrared finiteness, and the last term in (2.20) does not contribute to logarithmic mass renormalization. Multiple insertions of the mass term to a single propagator in any diagram or a mass insertion to any propagator in logarithmically divergent diagrams do not contribute to mass renormalization\footnote{The parameter $\mu$ is introduced after the evaluation of Feynman diagrams in ordinary formulation. The scheme (2.20) is useful to see that the logarithmic renormalization of $m_{0}^{2}$ is sufficient for mass renormalization and that no renormalization of the parameter $\mu^{2}$ takes place.}. The logarithmically divergent diagrams (or sub-diagrams)  are handled by the propagator (2.19) without encountering the infrared divergence. In fact, this is a neater way to see how the definition of mass independent renormalization factors works. We thus recognize that the 
crucial ingredient of our analysis is the elimination of all the quadratic divergences by the counter term $\Delta_{sub}(\lambda_{0},M^{2})$.

We here get contact with the treatment of truly {\em massless} $\lambda\phi^{4}$ theory by Zinn-Justin~\cite{zinn-justin} if one sets 
$m_{0}^{2}=0$ in (2.18). Our analysis indicates that we can eliminate all the ultraviolet divergences in massless $\lambda\phi^{4}$ theory up to any finite orders of the mass insertion term in (2.18) without any mass renormalization, but the difference from the massive theory is that one recovers the massless propagator, which is generally plagued with infrared divergences, when one sums all the mass insertion terms (so-called "spring diagrams"). The treatment of the truly massless $\lambda\phi^{4}$ theory by avoiding exceptional momenta~\cite{zinn-justin} is beyond the scope of the present analysis.     

\section{Inhomogeneous renormalization group equation}  

\subsection{Renormalization scale in subtractive renormalization}

We now examine if a more general class of theories are defined for the starting Lagrangian (2.1).
From the comparison of (2.1) with  (2.3) or  (2.6), one sees that  a different theory appears depending on the different choice of the subtraction term $\Delta_{sub}$ which is introduced  simultaneously with a specific regularization. The term $\Delta_{sub}$ is usually chosen to subtract all the quadratic divergences to define a finite theory, but one generally has more freedom in the choice of $\Delta_{sub}$ when one supposes that the magnitude of a large fixed cut-off $M$ has some physical significance\footnote{This extra freedom in $\Delta_{sub}$ may not be unnatural if one remembers that  $\lambda\phi^{4}$ theory does not belong to a "strictly renormalizable" theory in the parlance of Weinberg\cite{weinberg}.}.  

We thus examine the possibility
\begin{eqnarray}
{\cal L}&=&-\frac{1}{2}\phi_{0}(x)[-\Box(\frac{-\Box+M^{2}}
{M^{2}})^{2}+m_{0}^{2}]\phi_{0}(x)-\frac{1}{4!}\lambda_{0}\phi_{0}(x)^{4}\nonumber
\\
&&+\frac{1}{2}\Delta_{sub}(\lambda_{0},M^{2}, \mu^{2})\phi_{0}(x)^{2}
\end{eqnarray}
where $\mu$ is the parameter which specifies the renormalization mass scale. The quadratic divergence implies that we generally have two  subtraction constants, and {\em we introduce  the  same renormalization point $\mu$ to the subtractive renormalization also}. One may also rewrite the main part of the Lagrangian (3.1) by using (2.18) to emphasize the use of the common $\mu$ for both subtractions. For a technical reason to simplify the mass insertion term in the analysis of renormalization group equations, we use the notation of (3.6) but work in {\em massive} perturbation theory. The $\mu$-dependence of $\Delta_{sub}(\lambda_{0},M^{2}, \mu^{2})$  implies that the physical mass depends on $\mu$ explicitly, a situation  unconventional in ordinary renormalization theory. Note that our replacement in (3.1) is meaningful only for the formulation with a fixed large cut-off $M$. It has no meaning in dimensional regularization, for example.

To be more specific, we use the definition 
\begin{eqnarray}
\Delta_{sub}(\lambda_{0},M^{2}, \mu^{2})=\Delta_{sub}(\lambda_{0},M^{2})
+\delta\Delta_{sub}(\lambda_{0},M^{2}, \mu^{2}),
\end{eqnarray}
and the term $\Delta_{sub}(\lambda_{0},M^{2})$ subtracts 
all the quadratic divergences before the operation of ordinary renormalization as in (2.3) or (2.6).
The extra term 
\begin{eqnarray}
\delta\Delta_{sub}(\lambda_{0},M^{2}, \mu^{2})=-\lambda_{0}\mu^{2}f(\lambda_{0}, \frac{M^{2}}{\mu^{2}})
\end{eqnarray}
which is proportional to $\lambda_{0}$ and $\mu^{2}$ gives rise to an extra induced mass~\footnote{If one chooses $\delta\Delta_{sub}(\lambda_{0},M^{2}, \hat{m}_{0}^{2})$ with a constant $\hat{m}_{0}^{2}$ which is independent of $\mu$, such a term may generally be absorbed into a redefinition of the bare mass 
$m_{0}^{2}$.}. The choice of this term is rather arbitrary and we choose 
$\delta\Delta_{sub}(\lambda_{0},M^{2}, \mu^{2})$ order by order in perturbation theory to make the extra induced term proportional to $\mu^{2}$ finite by a suitable renormalization; the ultraviolet divergence in $\delta\Delta_{sub}(\lambda_{0},M^{2}, \mu^{2})$ is at most logarithmic as is the case of the ordinary mass insertion term. 
It is important to subtract the divergence proportional to $\mu^{2}$
by a term contained in $\delta\Delta_{sub}(\lambda_{0},M^{2}, \mu^{2})$,          
 which is higher order in $\lambda_{0}$. We thus avoid the operator 
mixing between $m_{0}^{2}\phi_{0}(x)^{2}$ and $\delta\Delta_{sub}(\lambda_{0},M^{2}, \mu^{2})\phi_{0}(x)^{2}$ through the logarithmic divergence.

This procedure does not interfere with the ordinary
multiplicative renormalization of $\phi_{0}$, $m_{0}^{2}$ and $\lambda_{0}$ in (2.10) which are at most logarithmically divergent (after the complete subtraction of quadratic divergences by $\Delta_{sub}(\lambda_{0},M^{2})$) in our modified  mass-independent scheme. 

\subsection{ Renormalization group equation}

To discuss the renormalization group equation, we start with the Feynman path integral 
\begin{eqnarray}
\langle T\phi_{0}(x_{1})... \phi_{0}(x_{n})\rangle=
\int{\cal D}\phi_{0}\phi_{0}(x_{1})... \phi_{0}(x_{n})\exp\{\int d^{4}x {\cal L}\}
\end{eqnarray}
where we retain only the connected components in $\langle T\phi_{0}(x_{1})... \phi_{0}(x_{n})\rangle$. By noting that our starting Lagrangian (3.1) depends on the parameter $\mu$ through the term $\Delta_{sub}(\lambda_{0},M^{2}, \mu^{2})$, we have 
\begin{eqnarray}
&&\mu\frac{d}{d\mu}\langle T\phi_{0}(x_{1})..... \phi_{0}(x_{n})\rangle
\nonumber\\
&&-\frac{1}{2}\left(\mu\frac{\partial}{\partial\mu}\delta\Delta_{sub}(\lambda_{0},M^{2}, \mu^{2})\right)\int d^{4}x\langle T\phi^{2}_{0}(x)\phi_{0}(x_{1})..... \phi_{0}(x_{n})\rangle=0, \nonumber\\
\end{eqnarray}
which corresponds to Schwinger's action principle in operator formalism and an {\em identity}. The appearance of the inhomogeneous renormalization group equation
(3.5) is not surprising for a theory which is subtractively renormalized~\cite{zee}.
This relation can also be written as 
\begin{eqnarray}
&&\mu\frac{d}{d\mu}\langle T\phi_{0}(x_{1})..... \phi_{0}(x_{n})\rangle
\nonumber\\
&&+\left(\mu\frac{\partial}{\partial\mu}\delta\Delta_{sub}(\lambda_{0},M^{2}, \mu^{2})\right)\frac{\partial}{\partial m_{0}^{2}}\langle T\phi_{0}(x_{1})..... \phi_{0}(x_{n})\rangle=0, \nonumber\\
\end{eqnarray}
in the present mass independent renormalization scheme. The derivative with respect to $m_{0}^{2}$ is taken with fixed $\lambda_{0}, \ M$ and $\mu$. 

In terms of the 1PI(single particle irreducible) vertex function, we have
\begin{eqnarray}
&&\mu\frac{d}{d\mu}\Gamma_{n(0)}(x_{1}, ...., x_{n})\nonumber\\
&&+\left(\mu\frac{\partial}{\partial\mu}\delta\Delta_{sub}(\lambda_{0},M^{2}, \mu^{2})\right)\frac{\partial}{\partial m_{0}^{2}}\Gamma_{n(0)}(x_{1}, ...., x_{n})=0,
\end{eqnarray}
or after renormalization
\begin{eqnarray}
&&(\sqrt{Z})^{n}\mu\frac{d}{d\mu}\left((\sqrt{Z})^{-n}\Gamma_{n}(x_{1}, ...., x_{n})\right)\nonumber\\
&&+\left(\mu\frac{\partial}{\partial\mu}\delta\Delta_{sub}(\lambda_{0},M^{2}, \mu^{2})\right)\frac{Z}{Z_{m}}\frac{\partial}{\partial m^{2}}\Gamma_{n}(x_{1}, ...., x_{n})=0
\end{eqnarray}
in the present mass independent renormalization scheme. In this last operation the mass independent renormalization is essential.

If translated into the Fourier transformed vertex function, this implies 
\begin{eqnarray}
&&\{\mu\frac{\partial}{\partial \mu}+\beta\frac{\partial}{\partial\lambda}-\gamma_{m}m^{2}\frac{\partial}{\partial m^{2}}-n\gamma_{\phi}\}\nonumber\\
&&\times\Gamma_{n}(p_{1}, ...., p_{n})
+\left(\mu\frac{\partial}{\partial\mu}\delta\Delta_{sub}(\lambda_{0},M^{2}, \mu^{2})\right)\frac{Z}{Z_{m}}\frac{\partial}{\partial m^{2}}\Gamma_{n}(p_{1}, ...., p_{n})=0\nonumber\\
\end{eqnarray}
which can also be written as 
\begin{eqnarray}
&&\{\mu\frac{\partial}{\partial \mu}+\beta\frac{\partial}{\partial\lambda}-\left(\gamma_{m}m^{2}+\gamma_{m}^{(2)}\mu^{2}\right)\frac{\partial}{\partial m^{2}}-n\gamma_{\phi}\}\Gamma_{n}(p_{1}, ...., p_{n})=0\nonumber\\
\end{eqnarray}
with
\begin{eqnarray}
\gamma_{m}^{(2)}\mu^{2}\equiv -\left(\mu\frac{\partial}{\partial\mu}\delta\Delta_{sub}(\lambda_{0},M^{2}, \mu^{2})\right)\frac{Z}{Z_{m}}.
\end{eqnarray}
In these equations (3.9) and (3.10), we defined renormalization group parameters by the standard manner 
\begin{eqnarray}
&&\gamma_{m}=-\left(\mu\frac{d}{d\mu}m^{2}(\mu)\right)/m^{2}=\left(\mu\frac{d}{d\mu}(\frac{Z_{m}}{Z})\right)(\frac{Z}{Z_{m}}),\nonumber\\
&&\beta=\mu\frac{d}{d\mu}\lambda(\mu),\nonumber\\
&&\gamma_{\phi}=\frac{1}{2}\frac{1}{Z}\mu\frac{d}{d\mu}Z
\end{eqnarray}
where the derivative is taken with fixed bare parameters and $M$.
From a dimensional analysis, we have 
$\delta\Delta_{sub}(\lambda_{0},M^{2}, \mu^{2})=-\lambda_{0}\mu^{2}f(\lambda_{0}, \frac{M^{2}}{\mu^{2}})$ 
with a suitable function $f(x,y)$ as in (3.3). If one wants to have a finite renormalization group equation, we need to have
$\gamma_{m}^{(2)}\mu^{2}=\lambda\mu^{2}P(\lambda)$ in (3.10)
where $P(x)$ is a suitable polynomial in $x$.
We note a relation
\begin{eqnarray}
\gamma_{m}^{(2)}(\lambda)\mu^{2}&=&\mu\frac{d}{d\mu}\left(\lambda_{0}\mu^{2}f(\lambda_{0}, \frac{M^{2}}{\mu^{2}})
\right)\frac{Z}{Z_{m}}\nonumber\\
&=&\mu\frac{d}{d\mu}\left(\lambda_{0}\mu^{2}f(\lambda_{0}, \frac{M^{2}}{\mu^{2}})\frac{Z}{Z_{m}}
\right)\nonumber\\
&&-\left(\lambda_{0}\mu^{2}f(\lambda_{0}, \frac{M^{2}}{\mu^{2}})
\frac{Z}{Z_{m}}\right)\mu\frac{d}{d\mu}(\frac{Z}{Z_{m}})\frac{Z_{m}}{Z}\nonumber\\
&=&\mu\frac{d}{d\mu}\left(\lambda_{0}\mu^{2}f(\lambda_{0}, \frac{M^{2}}{\mu^{2}})\frac{Z}{Z_{m}}
\right)\nonumber\\
&&+\left(\lambda_{0}\mu^{2}f(\lambda_{0}, \frac{M^{2}}{\mu^{2}})
\frac{Z}{Z_{m}}\right)\gamma_{m}
\end{eqnarray}
where the derivative is taken with fixed bare parameters and $M$. This is also written as 
\begin{eqnarray}
\gamma_{m}^{(2)}\mu^{2}=\mu\frac{d}{d\mu}\hat{m}^{2}+\gamma_{m}\hat{m}^{2}
\end{eqnarray}
with the {\em induced renormalized mass} 
\begin{eqnarray}
\hat{m}^{2}=\left(\lambda_{0}f(\lambda_{0}, \frac{M^{2}}{\mu^{2}})
\frac{Z}{Z_{m}}\right)\mu^{2}.
\end{eqnarray}
The finite induced renormalized mass implies the finite renormalization group equation and vise versa, as it should be. 

The simplest and explicit example of such an induced renormalized mass is given by the choice $f(\lambda_{0}, \frac{M^{2}}{\mu^{2}})=c_{1}Z Z_{m}/Z_{\lambda}$ with a numerical constant $c_{1}$, and the finite induced renormalized mass 
 $\hat{m}^{2}=\left(\lambda_{0}f(\lambda_{0}, \frac{M^{2}}{\mu^{2}})\frac{Z}{Z_{m}}\right)\mu^{2}\\=c_{1}\lambda\mu^{2}$. In general, one can choose 
\begin{eqnarray}
\hat{m}^{2}=\mu^{2}\sum_{k=1}^{\infty}c_{k}\lambda^{k}=\mu^{2}g(\lambda)
\end{eqnarray}
with (arbitrary) numerical constants $c_{1}, c_{2}, .....$, and 
\begin{eqnarray}
\gamma_{m}^{(2)}\mu^{2}&=&(2+\beta\frac{d}{d\lambda} \ln g(\lambda)+\gamma_{m})\hat{m}^{2}
\end{eqnarray}
from (3.14).
The renormalization group equation (3.10) is thus written as 
\begin{eqnarray}
\{\mu\frac{\partial}{\partial \mu}+\beta\frac{\partial}{\partial\lambda}-\left(\gamma_{m}m^{2}+\hat{\gamma}_{m}\hat{m}^{2}\right)\frac{\partial}{\partial m^{2}}-n\gamma_{\phi}\}\Gamma_{n}(p_{1}, ...., p_{n})=0
\end{eqnarray}
with 
\begin{eqnarray}
\hat{\gamma}_{m}\equiv2+\beta\frac{d}{d\lambda} \ln g(\lambda)+\gamma_{m}.
\end{eqnarray}
The appearance of the term with $\hat{m}^{2}$ is a new feature of our equation. The mass $\hat{m}^{2}$ in (3.15) plays two roles in the renormalization group equation (3.18) ; firstly as a part of the renormalized mass
\begin{eqnarray}
m^{2}+\hat{m}^{2},
\end{eqnarray}
and secondly it responds to the operation $\mu\frac{\partial}{\partial \mu}$ and $\beta\frac{\partial}{\partial\lambda}$ through its explicit dependence on $\mu$ and $\lambda$.
Note that the bare Lagrangian (3.1) is written in terms of renormalized quantities as 
\begin{eqnarray}
{\cal L}&=&-\frac{Z}{2}\phi(x)[-\Box(\frac{-\Box+M^{2}}
{M^{2}})^{2}+\frac{Z_{m}}{Z}m^{2}]\phi(x)-\frac{Z_{\lambda}}{4!}\lambda\phi(x)^{4}\nonumber
\\
&&-\frac{1}{2}Z_{m}\hat{m}^{2}\phi(x)^{2}
+\frac{1}{2}Z\Delta_{sub}(\lambda_{0},M^{2})\phi(x)^{2}.
\end{eqnarray}

To study the scaling behavior of the proper vertex with respect to the 
scaling of momenta, it is convenient to treat $\hat{m}^{2}$ as if it does
not explicitly depend on $\mu$ nor on $\lambda$. This is achieved by replacing 
\begin{eqnarray}
\hat{\gamma}_{m}\rightarrow \hat{\gamma}^{\prime}_{m}\equiv \hat{\gamma}_{m}-2-\beta\frac{d}{d\lambda} \ln g(\lambda)=\gamma_{m},
\end{eqnarray}
and we have the renormalization group equation, instead of (3.18),
\begin{eqnarray}
\{\mu\frac{\partial}{\partial \mu}+\beta\frac{\partial}{\partial\lambda}-\left(\gamma_{m}m^{2}+\gamma_{m}\hat{m}^{2}\right)\frac{\partial}{\partial m^{2}}-n\gamma_{\phi}\}\Gamma_{n}(p_{1}, ...., p_{n})=0
\end{eqnarray}
where the mass $\hat{m}^{2}$ appearing in $\Gamma_{n}(p_{1}, ...., p_{n})$ is now treated as if it has no explicit $\mu^{2}$ nor $\lambda$ dependence. 
The solution of this last form of the renormalization group equation (3.23), which is essentially the same as the ordinary renormalization group equation, is written as  
\begin{eqnarray}
&&\Gamma_{n}(p_{1},..,p_{n}; \lambda, m^{2}+\hat{m}^{2},\mu^{2})
\nonumber\\
&=&\exp[-\int_{0}^{t}dt n\gamma_{\phi}(t)dt]\Gamma_{n}(p_{1},..,p_{n}; \lambda(t), m^{2}(t)e^{2t}+\hat{m}^{2}(t)e^{2t},\mu^{2}e^{2t})\nonumber\\
\end{eqnarray}
where $\gamma_{\phi}(t)=\gamma_{\phi}(\lambda(t))$ and 
\begin{eqnarray}
&&\frac{d}{dt}\lambda(t)=\beta(\lambda(t)), \hspace{1cm} \lambda(0)=\lambda,\nonumber\\
&&\frac{d}{dt}m^{2}(t)=-(2+\gamma_{m}(\lambda(t)))m^{2}(t), \hspace{1cm} m^{2}(0)=m^{2},\nonumber\\
&&\frac{d}{dt}\hat{m}^{2}(t)=-(2+\gamma_{m}(\lambda(t)))\hat{m}^{2}(t), \hspace{1cm} \hat{m}^{2}(0)=\hat{m}^{2}.
\end{eqnarray}
Note that the renormalization group running of $\hat{m}^{2}(t)$ is defined by $\gamma_{m}$. The relation (3.24) shows that both-hand sides vanish for the same set of momenta, as it should be. The relation (3.24) is also written in the form
\begin{eqnarray}
&&\Gamma_{n}(e^{t}p_{1},..,e^{t}p_{n}; \lambda, m^{2}+\hat{m}^{2},\mu^{2})
\nonumber\\
&=&\exp[nt-n\int_{0}^{t}dt \gamma_{\phi}(t)dt]\Gamma_{n}(p_{1},..,p_{n}; \lambda(t), m^{2}(t)+\hat{m}^{2}(t),\mu^{2})\nonumber\\
\end{eqnarray}
by taking the dimensional analysis of $\Gamma_{n}$ into account.
The scaling behavior of the proper vertex with respect to the uniform scaling of momenta, which is {\em defined} by the left-hand side of (3.26), is thus essentially the same as the conventional formula except for the appearance of $m^{2}(t)+\hat{m}^{2}(t)$. 

\subsection{Comparison with other renormalization group equations}

We here compare our renormalization group equation (3.10) or (3.18) with other known forms of the renormalization group equation. The conventional renormalization group equation is based on the Lagrangian (2.18) which is invariant under the variation of $\mu$ and thus one obtains the homogeneous equation without the extra term in (3.5). In the case of the Callan-Symanzik equation, one compares two different theories with masses $m_{0}^{2}$ and $m_{0}^{2}+\epsilon m_{0}^{2}$ with an infinitesimal parameter $\epsilon$ in the Lagrangian (2.3). In the order linear in $\epsilon$, one thus obtains by 
means of the action principle
\begin{eqnarray}
&&m_{0}\frac{d}{dm_{0}}
\langle T\phi_{0}(x_{1})..... \phi_{0}(x_{n})\rangle
\nonumber\\
&&+\int d^{4}x m_{0}^{2}\langle T[\phi^{2}]_{0}(x)\phi_{0}(x_{1})..... \phi_{0}(x_{n})\rangle=0
\end{eqnarray}
with
\begin{eqnarray}
[\phi^{2}_{0}](x)\equiv\phi_{0}(x)(\frac{-\Box+M^{2}}
{M^{2}})^{2}\phi_{0}(x),
\end{eqnarray}
which gives rise to the inhomogeneous Callan-Symanzik equation when one considers the renormalized Green's function with the renormalization factors in (2.2).  We also compare two different theories with $\mu$ and $\mu+\delta\mu$ in the derivation of (3.5) and this leads to an inhomogeneous equation analogous to the Callan-Symanzik equation. It is however interesting that we eventually obtain a homogeneous equation since we have an extra freedom $\mu$. 

It is obvious that our renormalization group equation has a meaning different from the conventional renormalization group equation only when we use the same parameter $\mu$ for both of the logarithmic and quadratic subtractions. If we assign separate constants to these two divergences, the extra term  $\delta\Delta_{sub}(\lambda_{0},M^{2}, \hat{m}_{0}^{2})$ with a constant $\hat{m}_{0}^{2}$ independent of $\mu$ in (3.2) is generally absorbed into a redefinition of the bare mass $m_{0}^{2}$, as we already noted. The conventional homogeneous renormalization group equation then holds with the modified bare mass.  This fact explains the scaling property of (3.26). 

From the point of view of the Callan-Symanzik equation, which is based on the variation of  $m_{0}^{2}$, our extra term  $\delta\Delta_{sub}(\lambda_{0},M^{2}, \mu^{2})$ in (3.1) is just another constant and is generally absorbed into a redefinition of the bare mass. The Callan-Symanzik equation, which can be treated by side-stepping quadratic divergences~\cite{callan}, is rather similar to the dimensional regularization; both are insensitive to the presence of quadratic divergences.

Our renormalization group equation has a  possible meaning in the formulation where  the large fixed cut-off of quadratic divergences plays an essential role and that some physical significance is attached to the magnitude of the large cut-off~\footnote{For a given bare Lagrangian with a specified regularization, any treatment gives essentially the same physics. Our renormalization group equation thus describes the same contents as the Callan-Symanzik equation with the modified bare mass, just as the conventional renormalization group equation and the Callan-Symanzik equation provide alternative descriptions in conventional formulation.}. In applications to particle and condensed matter physics, one often encounters such situations. 
Our scheme describes a continuous set of theories with different renormalized masses parameterized by $\mu$ for a large fixed 
cut-off $M$; for a large fixed $M$, different $\mu$ defines a different theory in (3.1). For each fixed $\mu$ we recover the conventional theory with a corresponding mass, although the $\mu$ independence of physical quantities is lost. By changing the parameter $\mu$, we interpolate between the theory defined by dimensional regularization for small $\mu$ and the theory with un-subtracted quadratic divergences for large $\mu\sim M$.
This picture, which derives a possible large mass by absorbing a part of quadratic divergences as higher order effects instead of enlarging the bare mass $m_{0}^{2}$ (which receives at most logarithmic divergences in our formulation), might have some relevance to the argument of "naturalness" in a theory with a large fixed cut-off $M$.
  
Our renormalization group equation (3.10) or (3.18) is similar to (1.1), but
their physical contents are different. Eq.(1.1) is proposed on the basis of an analysis of the 
invariance property of the Green's function under the change of $\Lambda$ in the Wilsonian renormalization~\cite{hughes}, while
our equation (3.18) is derived from the action principle (3.5) which states that the Green's function is not invariant under the change of $\mu$. Nevertheless, the parameter $\mu$ in our equation plays a role similar to the cut-off $\Lambda$ in (1.1), and for small $\mu\ll m$ and small $\Lambda\ll m$, both of our equation and (1.1) approach the conventional renormalization group equation.
 
\section{Simple example}

Coming back to the explicit example of the $\lambda\phi^{4}$ theory defined by (3.1), 
the one-loop mass correction in Fig.1a in the present higher derivative 
regularization is given by 
\begin{eqnarray}
&&\frac{\lambda_{0}}{2}\int_{-\infty}^{\infty}\frac{d^{4}k}{(2\pi)^{4}}\frac{1}{k^{2}+m_{0}^{2}}(\frac{M^{2}}{k^{2}+M^{2}})^{2}\nonumber\\
&=&\frac{\lambda_{0}}{32\pi^{2}}\int_{0}^{\infty}dk^{2}\frac{k^{2}}{k^{2}}(\frac{M^{2}}{k^{2}+M^{2}})^{2}
-\frac{\lambda_{0}m_{0}^{2}}{32\pi^{4}}\int_{-\infty}^{\infty}d^{4}k\frac{1}{(k^{2}+\mu^{2})^{2}}(\frac{M^{2}}{k^{2}+M^{2}})^{2}\nonumber\\
&&+\frac{\lambda_{0}m_{0}^{2}}{32\pi^{4}}\int_{-\infty}^{\infty}d^{4}k\frac{1}{k^{2}}[-\frac{\mu^{2}}{(k^{2}+\mu^{2})^{2}}
+\frac{(m_{0}^{2}-\mu^{2})}{(k^{2}+\mu^{2})(k^{2}+m_{0}^{2})}](\frac{M^{2}}{k^{2}+M^{2}})^{2}\nonumber\\
\end{eqnarray}
where we used the prescription in (2.13) to convert the massive perturbation theory to the mass independent scheme.
The first quadratically divergent term, which is independent of $m_{0}^{2}$, is infrared finite and gives 
\begin{eqnarray}
\frac{\lambda_{0}}{32\pi^{2}}\int_{0}^{\infty}dk^{2}\frac{k^{2}}{k^{2}}(\frac{M^{2}}{k^{2}+M^{2}})^{2}=\frac{\lambda_{0}}{32\pi^{2}}M^{2}
\end{eqnarray}
and the second logarithmically divergent term gives for large $M^{2}$
\begin{eqnarray}
-\frac{\lambda_{0}m_{0}^{2}}{32\pi^{4}}\int_{-\infty}^{\infty}d^{4}k\frac{1}{(k^{2}+\mu^{2})^{2}}(\frac{M^{2}}{k^{2}+M^{2}})^{2}=-\frac{\lambda_{0}m_{0}^{2}}{32\pi^{2}}\ln(\frac{M^{2}}{\mu^{2}}).
\end{eqnarray}
This defines the mass renormalization factor in the mass independent way. The third finite term gives
\begin{eqnarray}
&&\frac{\lambda_{0}m_{0}^{2}}{32\pi^{4}}\int_{-\infty}^{\infty}d^{4}k\frac{1}{k^{2}}[-\frac{\mu^{2}}{(k^{2}+\mu^{2})^{2}}
+\frac{(m_{0}^{2}-\mu^{2})}{(k^{2}+\mu^{2})(k^{2}+m_{0}^{2})}](\frac{M^{2}}{k^{2}+M^{2}})^{2}
\nonumber\\
&&=-\frac{\lambda_{0}m_{0}^{2}}{32\pi^{2}}[\ln\frac{\mu^{2}}{m_{0}^{2}}+1]
\end{eqnarray}
for large $M^{2}$. The other renormalization factors are $Z=1$ and $Z_{\lambda}=1$ to this order.
  
The two-point vertex function to this order is then given in the present scheme
\begin{eqnarray}
\Gamma_{2}(k)&=&k^{2}+m_{0}^{2}+\frac{\lambda_{0}}{32\pi^{2}}M^{2}-\frac{\lambda_{0}}{32\pi^{2}}m_{0}^{2}
\ln\frac{M^{2}}{\mu^{2}}-\frac{\lambda_{0}m_{0}^{2}}{32\pi^{2}}[\ln\frac{\mu^{2}}{m_{0}^{2}}+1]\nonumber\\
&&
-\Delta_{sub}(\lambda_{0},M^{2})
-\delta\Delta_{sub}(\lambda_{0},M^{2}, \mu^{2})\nonumber\\
&=&k^{2}+m^{2}\left(1-\frac{\lambda}{32\pi^{2}}(\ln\frac{\mu^{2}}{m^{2}}+1)\right)+\frac{\lambda}{32\pi^{2}}\mu^{2}+
O(\lambda_{0}^{2})
\end{eqnarray}
where we chose the renormalization factors to this order
\begin{eqnarray}
\Delta_{sub}(\lambda_{0},M^{2})=\frac{\lambda_{0}}{32\pi^{2}}M^{2}
\end{eqnarray}
and  
\begin{eqnarray}
m_{0}^{2}&=&Z_{m}m^{2},\nonumber\\
Z_{m}&=&1+\frac{\lambda_{0}}{32\pi^{2}}\ln\frac{M^{2}}{\mu^{2}}.
\end{eqnarray}
For an illustration, we chose the induced mass term in (4.5) to this order at 
\begin{eqnarray}
-\delta\Delta_{sub}(\lambda_{0},M^{2}, \mu^{2})=\lambda_{0}\mu^{2}f(\lambda_{0}, \frac{M^{2}}{\mu^{2}})=\frac{\lambda_{0}}{32\pi^{2}}\mu^{2}.
\end{eqnarray}
The renormalization group parameters are then given to this order by 
$\beta=0$ and $\gamma_{\phi}=0$, and 
\begin{eqnarray}
\gamma_{m}&=&\left(\mu\frac{\partial}{\partial\mu}(\frac{Z_{m}}{Z})\right)(\frac{Z}{Z_{m}})=-\frac{\lambda}{16\pi^{2}}\nonumber\\
\gamma_{m}^{(2)}\mu^{2}&=&\mu\frac{\partial}{\partial\mu}\left(\lambda_{0}\mu^{2}f(\lambda_{0}, \frac{M^{2}}{\mu^{2}})
\right)\frac{Z}{Z_{m}}=\frac{\lambda}{16\pi^{2}}\mu^{2},
\end{eqnarray}
for a specific choice of (4.8).
One can then confirm that the renormalization group equation in (3.10) 
or equivalently (3.18)
\begin{eqnarray}
&&\{\mu\frac{\partial}{\partial \mu}-\left(\gamma_{m}m^{2}+\gamma_{m}^{(2)}\mu^{2}\right)\frac{\partial}{\partial m^{2}}\}\Gamma_{2}(k)=0
\end{eqnarray}
is satisfied by the above two-point proper vertex. 

The two-point proper vertex
\begin{eqnarray}
\Gamma_{2}(k)
&=&k^{2}+m^{2}\left(1-\frac{\lambda}{32\pi^{2}}(\ln\frac{\mu^{2}}{m^{2}}+1)\right)+\frac{\lambda}{32\pi^{2}}\mu^{2}
\end{eqnarray}
shows that the physical mass depends on the renormalization mass scale $\mu$. Note that the mass appearing in (4.11) stands for the physical mass to this leading order in perturbation theory since we have no wave function nor coupling constant renormalization to this order. The $\mu$-dependence of the renormalized mass $m$ is specified 
by the multiplicative renormalization factor $Z_{m}$ in $m_{0}^{2}=Z_{m}m^{2}$.
The second term in (4.11)
\begin{eqnarray}
m^{2}\left(1-\frac{\lambda}{32\pi^{2}}(\ln\frac{\mu^{2}}{m^{2}}+1)\right)\simeq m^{2}(\mu)\exp\left(-\frac{\lambda}{32\pi^{2}}(\ln\frac{\mu^{2}}{m^{2}}+1)\right)
\end{eqnarray}
is $\mu$-independent since the bare mass
\begin{eqnarray}
m_{0}^{2}&=&m^{2}Z_{m}\nonumber\\
&\simeq& 
m^{2}(\mu)\exp\left(\frac{\lambda}{32\pi^{2}}\ln\frac{M^{2}}{\mu^{2}}\right)
\end{eqnarray}
has the same $\mu$-dependence. The last term in (4.11) thus gives the 
renormalization scale $\mu$ dependence of the physical mass. This fact by itself is not surprising since our theory is formulated such that $\mu$ parameterizes a continuous set of theories with different physical masses. For a given physical mass the allowed range of $\mu$ is generally restricted, or else the physical mass is forced to be large by a large $\mu$ analogously to un-subtracted theory. In fact, if one sets $\mu \sim M$ one comes close to the result of un-subtracted theory (2.1). 
 The formula (4.11) agrees with the result of the dimensional regularization if the last term is set to zero.\footnote{A very small $\mu$ in our formulation gives rise to the conventional theory.} Our scheme thus
 interpolates between two different theories, namely, the theory defined by dimensional regularization and the theory with un-subtracted quadratic divergences; in the latter case the magnitude of a fixed large $M$ is supposed to have some physical significance.

The induced mass proportional to $\mu^{2}$ in the above simplest choice (4.11) is obtained by the replacement 
$M^{2}\rightarrow \mu^{2}$ in the quadratically divergent mass term, namely, the separation
\begin{eqnarray}
M^{2}=(M^{2}-\mu^{2})+\mu^{2}
\end{eqnarray}
or
\begin{eqnarray}
\mu^{2}=M^{2}-(M^{2}-\mu^{2})
\end{eqnarray}
where the first $M^{2}$ on the right-hand side arises from the Feynman diagram and $-(M^{2}-\mu^{2})$ arises from the counter term. This 
 is analogous to the separation of the logarithmic divergence
\begin{eqnarray}
\ln\frac{M^{2}}{m^{2}}&=&\ln\frac{M^{2}}{\mu^{2}}+\ln\frac{\mu^{2}}{m^{2}}\nonumber\\
&=&(\ln\frac{M^{2}}{m^{2}}-\ln\frac{\mu^{2}}{m^{2}})+\ln\frac{\mu^{2}}{m^{2}}
\end{eqnarray}
into divergent and finite parts. But this analogy does not work in general;
it is well-known that the signature of the quadratic divergence is 
opposite in bosonic and fermionic loop diagrams, and thus the positive 
coefficient of the quadratic divergent scalar mass is not guaranteed in general. The scheme proposed in the present paper is more general and it works for all the cases.

\section{Discussion and conclusion}

We have studied two aspects of quadratic divergences in this paper. The first is the mass independent renormalization scheme for a scalar theory with the higher derivative regularization, and the second is a possible subtractive renormalization with an inhomogeneous renormalization group equation.
  
We have argued that the mass independent scheme for a massive scalar theory is possible on the basis of the specific ansatz in the Lagrangain (2.3) or (2.18), which is supported by lower order calculations in perturbation theory.  This is consistent with previous analyses~\cite{zinn-justin}. The quadratic divergence depends only on the cut-off mass $M^{2}$ and $\lambda_{0}$ and thus it is quite "kinematical". This implies that  one can maintain the main physical contents, such as unitarity and analyticity, in tact by simply side-stepping the quadratic divergence in the dimensional regularization~\cite{t hooft} or by the use of the mass insertion technique in the Callan-Symanzik equation~\cite{callan}. The classical scaling argument of Bardeen~\cite{bardeen} may also be counted in this category. In fact, both of the Callan-Symanzik equation and the scaling argument of Bardeen are related to the quantum breaking of conformal symmetry, namely, the conformal anomaly.
Considering the  "kinematical" nature of the quadratic divergence, one may regard that the subtraction of the quadratic divergence in (2.3)
 or by the dimensional regularization is physically natural.

Alternatively, one may adopt a view that  the starting Lagrangian (2.1) allows a more freedom in the specification of subtractive renormalization. In this point of view, we discussed the possible use of a wider class of counter terms by allowing the appearance of the renormalization scale $\mu$ in the counter term of quadratic divergences;  to make this analysis sensible, we suppose that the magnitude of a large fixed cut-off $M$ has some physical significance. This scheme defines a continuous set of theories with different physical masses parameterized by $\mu$, unlike the conventional renormalization group which was introduced as a symmetry in the ordinary formulation such as in (2.18) and thus physical quantities are independent of the renormalization mass scale $\mu$. Our renormalization group equation is also similar to (1.1) proposed for the Wilsonian renormalization~\cite{hughes}, and the parameter $\mu$ plays a role similar, but not identical, to that of the cut-off $\Lambda$.  

A possible subtractive renormalization scheme we discussed, which interpolates between the theory with dimensional regularization and the theory with un-subtracted quadratic divergences, is unconventional in the framework of ordinary renormalization theory. Further analyses are required to see the physical relevance of such a scheme, but it may be useful in analyzing the broad aspects of quadratic divergences such as "naturalness" for a theory defined by a large fixed cut-off $M$.
\\

A very preliminary version of the present work was presented at Summer Institute 2009,
Fuji-Yoshida in Japan. I thank the participants of SI2009 for stimulating
discussions. I also thank H. Sonoda for useful correspondences and M. Ge and T. Inami for helpful comments.

\appendix

\section{$m_{0}^{2}$-independence of the quadratic divergence}

We illustrate the calculation of the simplest two-loop self-energy correction in $\lambda\varphi^{4}$ theory defined by (2.3)  (see Fig.1b) but without the higher derivative regularization for a moment. To analyze the quadratic divergence, it is sufficient to analyze the case with vanishing external momentum. We thus examine the integral of the form
\begin{eqnarray}
\int d^{4}kd^{4}l\frac{1}{k^{2}+m_{0}^{2}}\frac{1}{(k+l)^{2}+m_{0}^{2}}\frac{1}{l^{2}+m_{0}^{2}}
\end{eqnarray}
by ignoring the coupling constant $\lambda_{0}^{2}$ and other numerical factors. 

We want to show that the quadratically divergent part of the above integral with a momentum cut-off at $M$ is infrared finite for $m_{0}^{2}=0$. We first evaluate
\begin{eqnarray}
&&\int d^{4}l\frac{1}{(k+l)^{2}+m_{0}^{2}}\frac{1}{l^{2}+m_{0}^{2}}
\nonumber\\
&=&\int_{0}^{1}d\alpha\int d^{4}l\frac{1}{[(l+\alpha k)^{2}+\alpha(1-\alpha)k^{2}+m_{0}^{2}]^{2}}\nonumber\\
&=&\pi^{2}\int_{0}^{1}d\alpha\int_{0}^{M^{2}} dl^{2}\frac{l^{2}}{[l^{2}+\alpha(1-\alpha)k^{2}+m_{0}^{2}]^{2}}
\end{eqnarray}
where $\alpha$ is the Feynman parameter. The integral (A.2) is evaluated 
as
\begin{eqnarray}
\pi^{2}\int_{0}^{1}d\alpha\{\ln \left(\frac{M^{2}+\alpha(1-\alpha)k^{2}+m_{0}^{2}}{\alpha(1-\alpha)k^{2}+m_{0}^{2}}\right)-\frac{M^{2}}{M^{2}+\alpha(1-\alpha)k^{2}+m_{0}^{2}}\}
\end{eqnarray}
We thus evaluate the following integrals in (A.1) by noting 
$d^{4}k=\pi^{2}M^{2}xdx$ with $k^{2}=M^{2}x$ 
\begin{eqnarray}
&&M^{2}\int_{0}^{1}d\alpha\int_{0}^{1}dx\frac{x}{x+\epsilon}\ln \left(\frac{\alpha(1-\alpha)x+\epsilon+1}{\alpha(1-\alpha)x+\epsilon}\right)\nonumber\\
&=&M^{2}\int_{0}^{1}d\alpha\int_{0}^{1}dx\ln \left(\frac{\alpha(1-\alpha)x+\epsilon+1}{\alpha(1-\alpha)x+\epsilon}\right)\nonumber\\
&&-m_{0}^{2}\int_{0}^{1}d\alpha\int_{0}^{1}dx\frac{1}{x+\epsilon}\ln \left(\frac{\alpha(1-\alpha)x+\epsilon+1}{\alpha(1-\alpha)x+\epsilon}\right)
\end{eqnarray}
and 
\begin{eqnarray}
&&M^{2}\int_{0}^{1}d\alpha\int_{0}^{1}dx\frac{x}{x+\epsilon}\frac{1}{\alpha(1-\alpha)x+1+\epsilon}\nonumber\\
&=&M^{2}\int_{0}^{1}d\alpha\int_{0}^{1}dx\frac{1}{\alpha(1-\alpha)x+1+\epsilon}\nonumber\\
&&-m_{0}^{2}\int_{0}^{1}d\alpha\int_{0}^{1}dx\frac{1}{x+\epsilon}\frac{1}{\alpha(1-\alpha)x+1+\epsilon}
\end{eqnarray}
with $\epsilon=m_{0}^{2}/M^{2}$. It is confirmed that the terms with $m_{0}^{2}$ are at most logarithmically divergent for $\epsilon\rightarrow 0$ and thus those terms do not give rise to any quadratic 
divergence. The quadratically divergent term in (A.4) gives
\begin{eqnarray}
&&M^{2}\int_{0}^{1}d\alpha\int_{0}^{1}dx\ln \left(\frac{\alpha(1-\alpha)x+\epsilon+1}{\alpha(1-\alpha)x+\epsilon}\right)\nonumber\\
&&=M^{2}\int_{0}^{1}d\alpha\int_{0}^{1}dx\ln \left(\frac{\alpha(1-\alpha)x+1}{\alpha(1-\alpha)x}\right)
\end{eqnarray}
for $\epsilon=0$, and the quadratically divergent term in (A.5) gives
\begin{eqnarray}
M^{2}\int_{0}^{1}d\alpha\int_{0}^{1}dx\frac{1}{\alpha(1-\alpha)x+1+\epsilon}=M^{2}\int_{0}^{1}d\alpha\int_{0}^{1}dx\frac{1}{\alpha(1-\alpha)x+1}
\end{eqnarray}
for $\epsilon=0$, both of which are (infrared) finite. This shows that there is no divergence of the form 
$M^{2}\ln(M^{2}/m_{0}^{2})$. We have of course a divergence
such as $m_{0}^{2}\ln(M^{2}/m_{0}^{2})$ which is logarithmic.
This analysis which is based on a simple momentum cut-off is extended to the regularized Lagrangian in (2.3) and also to the mass independent scheme in (2.6).

More directly, on the basis of power counting argument one can confirm 
the infra-red finiteness of 
\begin{eqnarray}
\int d^{4}kd^{4}l\frac{1}{k^{2}}\frac{1}{(k+l)^{2}}\frac{1}{l^{2}}
\end{eqnarray}
which is obtained from (A.1) by setting $m^{2}_{0}=0$. One may then study
\begin{eqnarray}
\int d^{4}kd^{4}l\left(\frac{1}{(k+p)^{2}}-\frac{1}{k^{2}}\right)\frac{1}{(k+l)^{2}}\frac{1}{l^{2}}
\end{eqnarray}
in the mass independent scheme such as in (2.9). Here $p$ is the external momentum. This integral is free of the quadratic divergence although logarithmically divergent and thus needs a regularization as in (2.9).


\begin{thebibliography}{99}
\bibitem{bardeen}
The scaling argument of W. Bardeen, which emphasizes the classical scale
invariance in the Standard Model when all the bare mass parameters are 
set to zero, may support the scheme of the dimensional regularization.
See,\\
W.A.  Bardeen, "On naturalness in the standard model", FERMILAB-CONF-95-391-T.\\ 
In this spirit, the success of the Standard Model combined with the dimensional regularization might even suggest that the dimension of our space-time is actually close to  $D=4-\epsilon$ with $\epsilon\rightarrow 0$.
\end{thebibliography}
\end{document}